\documentclass{article}

\PassOptionsToPackage{numbers, compress}{natbib}

\usepackage{amsthm}
\usepackage{amssymb}
\usepackage[preprint]{neurips_2023}

\usepackage[utf8]{inputenc}
\usepackage[T1]{fontenc}
\usepackage{hyperref}
\usepackage{url}
\usepackage{booktabs}
\usepackage{amsfonts}
\usepackage{nicefrac}
\usepackage{microtype}
\usepackage{xcolor}
\usepackage{graphicx}
\usepackage{amsmath}
\usepackage{float}
\usepackage{listings}
\usepackage{enumitem}
\usepackage{array}
\usepackage{tabularx}
\usepackage{multirow}
\usepackage{tikz}
\usetikzlibrary{arrows.meta, shapes.geometric, positioning, fit, backgrounds, calc}
\usepackage{pgfplots}
\pgfplotsset{compat=1.18}

\hypersetup{
    colorlinks=false,
    pdfborder={0 0 0},
}

\title{AI Assurance: A Comprehensive Testing Strategy\\for Enterprise AI Systems}

\author{
  Chitra Badagi \\
  Thoughtworks Technologies \\
  Pune, MH, India \\
  \texttt{chitra.badagi@thoughtworks.com}
  \And
  Divye Singh \\
  Thoughtworks Technologies \\
  Pune, MH, India \\
  \texttt{divye.singh@thoughtworks.com}
  \And
  Animesh Sen \\
  Thoughtworks Technologies \\
  Hyderabad, TG, India \\
  \texttt{animesh.sen@thoughtworks.com}
  \And
  Adinath Shirsath \\
  Thoughtworks Technologies \\
  Pune, MH, India \\
  \texttt{adinaths@thoughtworks.com}
}

\begin{document}

\maketitle

\begin{abstract}
Enterprise AI systems, built on large language models, retrieval pipelines and autonomous agents, introduce a class of risks that traditional software quality assurance was never designed to address. These systems are probabilistic, context-sensitive and emergent: they cannot be verified to be correct in the classical sense, but only evaluated with increasing confidence.
This paper presents a comprehensive assurance strategy for enterprise AI systems built around three key principles: first, that AI testing should focus on continuous risk reduction rather than strict correctness verification; second, that evaluation must be treated as a core engineering discipline alongside development; and third, that failures in AI assurance can lead to organizational impacts that are fundamentally different from those seen in traditional deterministic software systems. We introduce a structured AI Failure Taxonomy, propose a revised five-layer AI Assurance Pyramid and provide operational guidance on evaluation-driven development, RAG system testing, model lifecycle management and governance. The goal is to equip engineering leaders and practitioners with a strategy that is both philosophically grounded and operationally deployable.

\end{abstract}

\clearpage


\tableofcontents

\clearpage
\section*{Summary}
\addcontentsline{toc}{section}{Summary}

Enterprise AI systems are failing in ways that traditional quality assurance was not built to detect. Not through crashes or error codes but through confident hallucinations, silent behavioural drift and coordination failures that produce wrong answers visually indistinguishable from correct ones. The teams most exposed are those that adopted AI capabilities quickly and evaluated them the same way they evaluate conventional software: with test suites, pass/fail gates and release-time validation. This approach is insufficient. It is not a resourcing problem. It is a structural mismatch between the assurance model and the system it is trying to verify.

\textbf{The core problem.} AI systems are probabilistic infrastructure. Their behaviour is not defined by code alone it is shaped by model training, prompt design, retrieval quality and the interactions between components. The same input can produce different outputs on different runs. A model update by a cloud provider can silently change behaviour across every prompt in the application. These properties make traditional verification impossible and make continuous evaluation mandatory.

\textbf{What this paper provides.} A comprehensive AI assurance strategy built on three
foundations: a philosophy of testing as continuous risk reduction (not correctness verification); a structured taxonomy of AI-native failure modes that makes test coverage systematic rather than intuitive; and a five-layer AI Assurance Pyramid that maps failure classes to the evaluation mechanisms designed to catch them.

\textbf{Key arguments.}
\begin{itemize}[leftmargin=*]
  \item Evaluation is not a QA activity layered onto development. In AI systems, evaluations define acceptable behaviour boundaries. Evaluation-Driven Development structuring the engineering loop around evaluation datasets is the natural consequence of working with systems that cannot be fully programmed.
  \item The AI testing pyramid must be redesigned. The original four-layer model does not reflect AI system architecture or testing economics. The revised five-layer pyramid (Layer 0 through Layer 4) separates deterministic infrastructure validation from probabilistic component, agent, multi-agent and outcome evaluations and maps each layer to specific failure classes.
  \item Model lifecycle management is a first-class engineering concern. Cloud-hosted model providers deprecate and update models on timelines outside developer's control. Prompt regression testing running continuously not just at release is the only reliable defence against silent behavioural degradation.
  \item Evaluation infrastructure must become a platform capability. Organisations running multiple AI systems cannot sustain per-project evaluation tooling. The evaluation stack, datasets, judges, rubrics, scoring pipelines, regression baselines must be built as shared infrastructure, not rebuilt for each deployment.
\end{itemize}

\textbf{For engineering leaders.} The investment case for evaluation infrastructure is a risk management case, not a quality engineering case. The cost of insufficient evaluation is measured in hallucination incidents, model drift discovered weeks after onset and adversarial failures that reach production. These costs consistently exceed the infrastructure investment they would have replaced.

\textbf{For engineering teams.} Prompts are behavioural specifications. They must be version-controlled, regression-tested and re-evaluated on every model change. Evaluation datasets are engineering artefacts with the same maintenance obligations as production code. Building and maintaining them is the skill that distinguishes teams that deploy AI reliably from those that deploy AI hopefully.


\clearpage

\section{Understanding Enterprise AI Systems}
\label{sec:landscape}

Enterprise AI systems span a spectrum of complexity from standalone language model
integrations to fully autonomous multi-agent workflows and each level introduces a distinct category of failure. Understanding this spectrum is the first step toward evaluating it.

\subsection{The AI Stack: From Language Models to Agentic Systems}

\paragraph{Large Language Models (LLMs).}
LLMs are trained on large text corpora to generate responses by predicting likely continuations. They do not retrieve answers from a fixed source, they construct them dynamically using statistical patterns learned during training. This has a direct consequence for testing: there is no single correct output to verify against. The same input can produce multiple semantically valid responses and the space of invalid responses is equally large and difficult to enumerate.

\textit{Analogy: An LLM is like an experienced consultant who synthesises an answer from memory and judgement. The answer may be insightful, partially wrong or confidently incorrect. The variability is not a bug, it is fundamental to how the system works. Testing must account for this.}

\paragraph{AI Agents.}
AI Agents extend LLMs with the ability to take action. An agent receives a goal, breaks it into steps, invokes tools APIs, databases, search engines, code interpreters and iterates until the goal is satisfied. This introduces multi-step reasoning chains, external dependencies and stateful execution all of which are independently capable of failure.

A critical nuance: an agent can produce a correct final answer via an incorrect reasoning path. This is a false positive and a production risk. Testing a single output tells the answer was right; only trajectory testing tells \textit{why} it was right and whether it will be right again.

\textit{Analogy: The consultant who can also act making calls, gathering data, booking things on your behalf. Now correctness depends not just on what they know, but what they do. A travel agent who books the right hotel at the wrong location has passed the output test but failed the task.}

\paragraph{Agentic AI Systems.}
Agentic AI represents multi-agent architectures where specialised agents collaborate, hand off context and make sequential decisions with minimal human intervention. At this level, system behaviour is not defined by any individual component; it \textit{emerges} from the interactions between them. A planning agent, a retrieval agent, a synthesis agent and a formatting agent may each behave correctly in isolation while producing unexpected outcomes in coordination.

This emergence property is what makes agentic systems qualitatively different from
microservices: there are no explicit contracts between agents. Communication is language-based and therefore subject to ambiguity, context loss and misinterpretation.

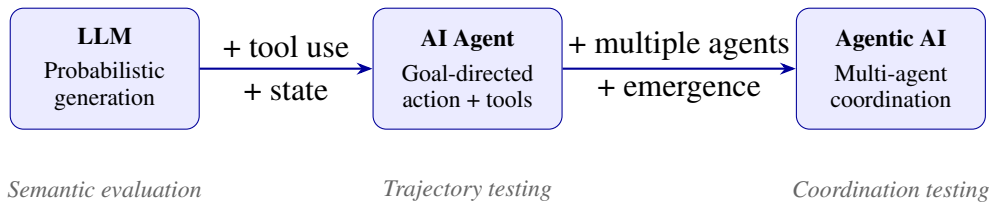
\begin{figure}[h]
\centering
\begin{tikzpicture}[
  box/.style={rectangle, rounded corners=5pt, draw=blue!60!black, fill=blue!8,
              minimum width=2.5cm, minimum height=1.6cm, align=center, font=\small},
  addbox/.style={rectangle, rounded corners=3pt, draw=orange!70!black, fill=orange!8,
                 minimum width=2.4cm, minimum height=0.75cm, align=center,
                 font=\footnotesize},
  teststyle/.style={font=\footnotesize\itshape, text=gray!80!black},
  arrow/.style={-{Stealth[length=6pt]}, thick, blue!60!black}
]
  \node[box] (llm) at (0,0) {\textbf{LLM}\\[3pt]\footnotesize Probabilistic\\generation};
  \node[box] (agent) at (4.8,0) {\textbf{AI Agent}\\[3pt]\footnotesize Goal-directed\\action + tools};
  \node[box] (agentic) at (10.4,0) {\textbf{Agentic AI}\\[3pt]\footnotesize Multi-agent\\coordination};

  \node[font=\large, text=black!60!black] (add1) at (2.4, 0.3) {+ tool use};
  \node[font=\large, text=black!60!black] (add2) at (2.4,-0.3) {+ state};
  \node[font=\large, text=black!60!black] (add3) at (7.6, 0.3) {+ multiple agents};
  \node[font=\large, text=black!60!black] (add4) at (7.6,-0.3) {+ emergence};

  \draw[arrow] (llm.east) -- (agent.west);
  \draw[arrow] (agent.east) -- (agentic.west);

  \node[teststyle, below=0.55cm of llm]  {Semantic evaluation};
  \node[teststyle, below=0.55cm of agent] {Trajectory testing};
  \node[teststyle, below=0.55cm of agentic] {Coordination testing};
\end{tikzpicture}
\caption{The AI system complexity spectrum. Each level adds new properties and demands
distinct testing approaches. Assurance strategies that work at the LLM level are
necessary but insufficient at the agentic level.}
\label{fig:ai-stack}
\end{figure}

\paragraph{The Evolution of Testing Complexity.}
The progression from LLM to Agent to Agentic AI mirrors how software evolved from monoliths to distributed systems. Each step increases capability and also introduces new categories of failure. Figure~\ref{fig:ai-stack} illustrates this spectrum; Table~\ref{tab:complexity}
summarises the primary testing challenges introduced at each level.

\begin{table}[h]
\centering
\caption{Testing complexity by AI system level}
\label{tab:complexity}
\begin{tabularx}{\linewidth}{@{} l l X @{}}
\toprule
\textbf{Level} & \textbf{Key Property} & \textbf{Primary Testing Challenge} \\
\midrule
LLM       & Probabilistic generation  & No fixed expected output; semantic evaluation required \\
AI Agent  & Goal-directed action      & Trajectory correctness; false positives via wrong path \\
Agentic AI & Multi-agent emergence    & Coordination failures; context loss; emergent misbehaviour \\
\bottomrule
\end{tabularx}
\end{table}

\subsection{Why Enterprises Build Their Own AI: The RAG Paradigm}

Public AI tools like ChatGPT are trained on publicly available data. An enterprise's competitive advantage often lives in its proprietary data, decades of domain expertise, internal research, regulatory knowledge, customer history and operational procedures. No public model has seen this data and fine-tuning alone is insufficient for systems that must reason over large, frequently updated document corpora.

The most widely adopted solution is \textbf{Retrieval-Augmented Generation (RAG)}: an architecture that separates knowledge storage from language generation. Rather than embedding all knowledge in model weights, RAG retrieves relevant documents at query time and provides them as context to the generation model. The model reasons over what it was given, not what it memorised.

\begin{figure}[htbp]
\centering
\begin{tikzpicture}[
  docbox/.style={rectangle, rounded corners=4pt, draw=orange!70, fill=orange!10,
                 minimum width=1.9cm, minimum height=0.7cm, align=center, font=\small},
  procbox/.style={rectangle, rounded corners=4pt, draw=blue!60, fill=blue!8,
                  minimum width=1.9cm, minimum height=0.7cm, align=center, font=\small},
  llmbox/.style={rectangle, rounded corners=4pt, draw=green!60!black, fill=green!8,
                 minimum width=1.9cm, minimum height=0.7cm, align=center, font=\small},
  storenode/.style={cylinder, shape border rotate=90, draw=blue!60, fill=blue!15,
                    minimum width=1.6cm, minimum height=0.8cm, align=center, font=\small},
  arrow/.style={-{Stealth[length=5pt]}, thick},
  dasharrow/.style={-{Stealth[length=7.5pt]}, thick, dashed, black!60},
  headlabel/.style={font=\small\bfseries, text=blue!70!black},
  errlabel/.style={font=\footnotesize, text=red!70!black}
]
  \node[headlabel] at (0, 2.5) {Offline Ingestion};
  \node[docbox]  (docs)   at (0,   1.7) {Documents};
  \node[procbox] (chunk)  at (2.5, 1.7) {Chunking};
  \node[procbox] (emb1)   at (5,   1.7) {Embedding};
  \node[storenode](store) at (7.5, 1.7) {Vector\\Store};

  \draw[arrow] (docs)  -- (chunk);
  \draw[arrow] (chunk) -- (emb1);
  \draw[arrow] (emb1)  -- (store);

  \node[headlabel] at (0, 0.6) {Query Time};
  \node[docbox]  (query)    at (0,   -0.2) {User Query};
  \node[procbox] (emb2)     at (2.5, -0.2) {Embedding};
  \node[procbox] (search)   at (5,   -0.2) {Vector\\Search};
  \node[llmbox]  (llm)      at (7.5, -0.2) {LLM};
  \node[docbox]  (response) at (10.2,-0.2) {Response};

  \draw[arrow] (query)  -- (emb2);
  \draw[arrow] (emb2)   -- (search);
  \draw[arrow] (search) -- (llm);
  \draw[arrow] (llm)    -- (response);

  \draw[dasharrow] (store.south) -- ++(0,-0.45) -| (search.north)
    node[near start, fill=white, inner sep=1pt,font=\footnotesize, text=black] {retrieve};

  \draw[errlabel] (3.8,-1.0) node {$\leftarrow$ Retrieval failure zone $\rightarrow$};
  \draw[errlabel] (8.9,-1.0) node {$\leftarrow$ Generation failure zone $\rightarrow$};
  \draw[red!50, dashed, very thick, rounded corners=3pt]
    (1.3,-0.7) rectangle (6.1, 0.2);
  \draw[red!50, dashed, very thick, rounded corners=3pt]
    (6.4,-0.7) rectangle (11.5, 0.2);
\end{tikzpicture}
\caption{RAG system architecture showing two independent failure surfaces. A correct final
response does not confirm that retrieval was accurate; testing each pipeline independently
is required to localise failures.}
\label{fig:rag-arch}
\end{figure}
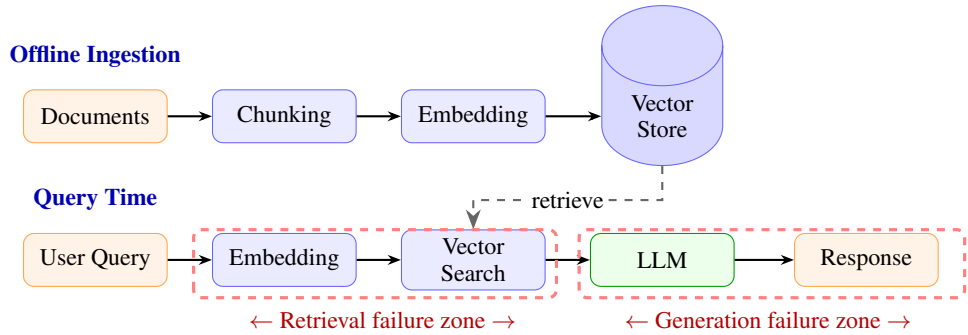

Figure~\ref{fig:rag-arch} shows the two-pipeline structure of a RAG system. This architecture introduces a critical testing constraint: \textbf{two independent components must work correctly} and they fail in different ways.

\begin{itemize}[leftmargin=*]
  \item \textbf{Retrieval} must surface the right documents for the query. Failure here manifests as responses that are coherently written but factually wrong or incomplete because the model was given the wrong evidence.
  \item \textbf{Generation} must reason faithfully over what was retrieved. Failure here manifests as responses that ignore, misinterpret or hallucinate beyond the provided context even when retrieval was accurate.
\end{itemize}

If only the final output is evaluated , it cannot be determined which component failed. A correct answer does not confirm correct retrieval. An incorrect answer may originate in retrieval, in generation or in both. This demands \textbf{independent evaluation of both components}: a requirement that fundamentally changes how RAG systems are tested.

Section~\ref{sec:rag-agentic} covers RAG evaluation strategy in detail, including the RAGAS
metrics framework designed specifically for this architecture.

\section{Why AI Systems Need a New Assurance Model}
\label{sec:assurance-case}

Traditional software quality assurance rests on a single foundational assumption: given the same input, the system will produce the same output. That assumption no longer holds. Every principle of conventional testing, fixed expected outputs, binary pass/fail verdicts, coverage as a completeness proxy is built on determinism. AI systems are probabilistic by design and this is not an edge case to work around. It is the central fact that demands a new model of assurance.

\subsection{The Determinism Problem}

In a traditional e-commerce checkout flow, the sequence of events is fixed. A user adds an item to cart, proceeds to payment and completes the purchase. The expected outcome is verifiable and binary: either the order was placed or it was not.

In an AI application, the same user input can produce multiple valid responses and also
multiple invalid ones that may appear just as plausible as the valid ones. There is no single expected output to assert against. The response space is effectively unbounded. This breaks the core abstraction of traditional testing at its foundation.

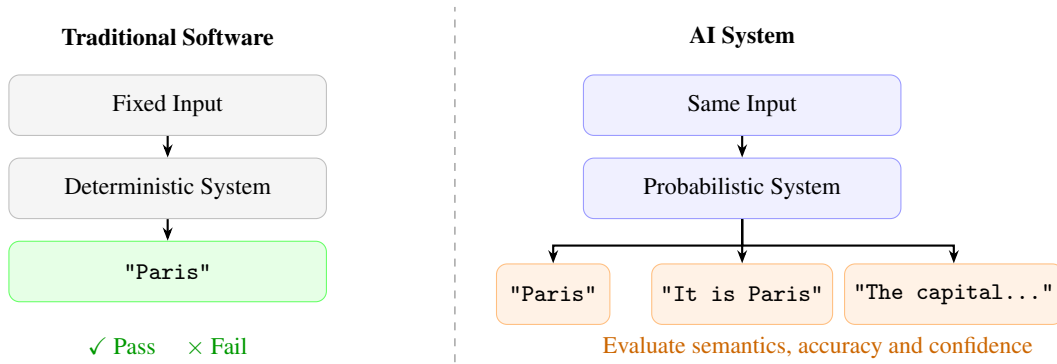
\begin{figure}[htbp]
\centering
\begin{tikzpicture}[
  box/.style={rectangle, rounded corners=4pt, draw, minimum width=4.2cm,
              minimum height=0.8cm, align=center, font=\small},
  probbox/.style={rectangle, rounded corners=4pt, draw, minimum width=4.2cm,
              minimum height=0.8cm, align=center, font=\small},
  arrow/.style={-{Stealth[length=5pt]}, thick},
  heading/.style={font=\small\bfseries}
]
  \node[heading]                              at (2.4, 4.0)  {Traditional Software};
  \node[box, fill=gray!8,   draw=gray!50]  (in1)  at (2.4, 3.1)  {Fixed Input};
  \node[box, fill=gray!8,   draw=gray!50]  (sys1) at (2.4, 2.0)  {Deterministic System};
  \node[box, fill=green!10, draw=green!60] (out1) at (2.4, 0.9)  {\texttt{"Paris"}};
  \node[font=\small, text=green!60!black]  (ver1) at (2.4, -0.1)
    {\checkmark~Pass \quad $\times$~Fail};

  \draw[arrow] (in1)  -- (sys1);
  \draw[arrow] (sys1) -- (out1);

  \draw[dashed, black!50] (6.2, -0.3) -- (6.2, 4.4);

  \node[heading]                               at (10.0, 4.0) {AI System};
  \node[box, fill=blue!6,   draw=blue!40]  (in2)  at (10.0, 3.1) {Same Input};
  \node[box, fill=blue!6,   draw=blue!40]  (sys2) at (10.0, 2.0) {Probabilistic System};

  \node[probbox, fill=orange!10, draw=orange!50, minimum width=1.5cm, align=center] (outA)
    at (7.5, 0.6) {\texttt{"Paris"}};
  \node[probbox, fill=orange!10, draw=orange!50, minimum width=1.5cm, align=center] (outB)
    at (10.0, 0.6) {\texttt{"It is Paris"}};
  \node[probbox, fill=orange!10, draw=orange!50, minimum width=1.5cm, align=center] (outC)
    at (12.8, 0.6) {\texttt{"The capital..."}};

  \draw[arrow] (sys2.south) -- ++(0,-0.35) -| (outA.north);
  \draw[arrow] (sys2.south) -- (outB.north);
  \draw[arrow] (sys2.south) -- ++(0,-0.35) -| (outC.north);

  \node[font=\small, text=orange!80!black] at (11.0, -0.1)
    {Evaluate semantics, accuracy and confidence};

  \draw[arrow] (in2) -- (sys2);
\end{tikzpicture}
\caption{Traditional testing relies on a single expected output and a binary verdict.
AI systems produce a distribution of valid responses that must be evaluated on meaning,
accuracy and appropriateness not literal string matching.}
\label{fig:determinism}
\end{figure}

Three properties compound this problem:

\begin{enumerate}[leftmargin=*]
  \item \textbf{Non-determinism.} The same prompt, submitted twice to the same model, may
    produce different outputs. Testing must shift from exact-match verification to semantic evaluation: does the response mean the right thing, not merely say the right words?

  \item \textbf{Context sensitivity.} LLMs are highly sensitive to phrasing ordering and framing. Minor prompt variations can produce qualitatively different outputs. The prompt itself becomes a test surface not just the system it invokes.

  \item \textbf{Emergent failure in composition.} Individual components can pass all unit-level tests while producing incorrect or unsafe behaviour in combination. A retrieval component that returns accurate documents and a generation model that produces grammatically sound responses can together produce a hallucinated answer. System-level evaluation is not optional.
\end{enumerate}

\subsection{Testing as Continuous Risk Reduction}

The correct mental model for AI assurance is not verification, it is \textbf{risk reduction}.

In deterministic systems, a passing test suite is a proof of correctness under the tested conditions. In AI systems, there is no such proof. The same evaluation run on the same system can return different scores on different days. A system that passes today may not pass tomorrow, not because anyone changed the code, but because the underlying model was quietly updated by the provider.

This demands a philosophical shift. The goal of testing AI systems is not to certify that the system is correct. It is to:

\begin{itemize}[leftmargin=*]
  \item \textbf{Reduce uncertainty}: narrow the range of behaviours the system is likely to exhibit
  \item \textbf{Bound failure}: establish known limits beyond which the system should not be deployed
  \item \textbf{Build confidence over time}: accumulate evidence that the system behaves acceptably across the distribution of inputs it will encounter in production
\end{itemize}

This reframing has a direct operational consequence: evaluation cannot be a release gate run once before deployment. It must run continuously, accumulate trends and trigger investigation when those trends degrade. A single passing run provides weak evidence. A consistent pattern of high scores across multiple inputs over weeks provides strong evidence.

\begin{figure}[htbp]
\centering
\begin{tikzpicture}[
  box/.style={rectangle, rounded corners=5pt, draw=blue!60!black, fill=blue!8,
              minimum width=2.5cm, minimum height=0.85cm, align=center, font=\small},
  gatebox/.style={rectangle, rounded corners=5pt, draw=green!60!black, fill=green!8,
                  minimum width=2.5cm, minimum height=0.85cm, align=center, font=\small},
  failbox/.style={rectangle, rounded corners=5pt, draw=red!60!black, fill=red!8,
                  minimum width=2.5cm, minimum height=0.85cm, align=center, font=\small},
  arrow/.style={-{Stealth[length=5pt]}, thick},
  dasharrow/.style={-{Stealth[length=7.5pt]}, thick, dashed, black!60},
  lbl/.style={font=\footnotesize, text=gray!80!black}
]
  \node[box]     (change)  at (0,   0)    {Code / Model\\/ Prompt Change};
  \node[box]     (eval)    at (3.5,   0)    {Evaluation\\Pipeline};
  \node[gatebox] (gate)    at (7,   0)    {Quality\\Gates};
  \node[gatebox] (deploy)  at (11,  0.9)  {Deploy};
  \node[failbox] (invest)  at (11, -0.9)  {Investigate};

  \draw[arrow] (change) -- (eval);
  \draw[arrow] (eval)   -- (gate);
  \draw[arrow] (gate.east) -- ++(0.4,0) |- (deploy.west)
    node[pos=0.55, fill=white, above, lbl] {pass};
  \draw[arrow] (gate.east) -- ++(0.4,0) |- (invest.west)
    node[pos=0.55, fill=white, below, lbl] {fail};

  \draw[dasharrow] (invest.south) -- ++(0,-0.5)
    -- ++(-11,0) node[midway, below, lbl] {prompt / config fix}
    -- (change.south);
\end{tikzpicture}
\caption{Evaluation infrastructure as the AI equivalent of CI/CD. Every change,
to code, model version or prompt triggers an evaluation pipeline with defined
quality gates. Unlike a test suite run once at release, this pipeline runs continuously.}
\label{fig:eval-cicd}
\end{figure}
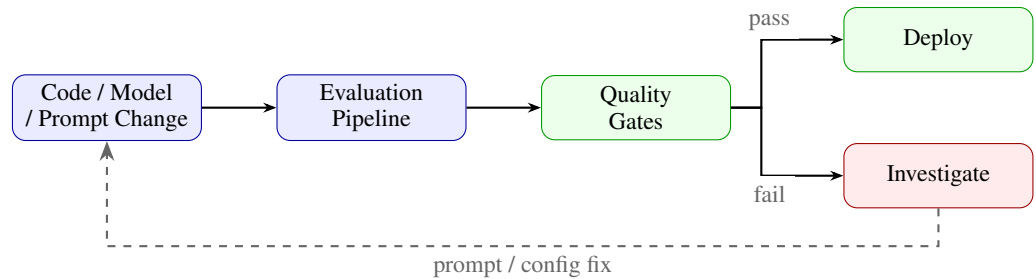

\textbf{The equivalent in traditional software is not the test suite, it is the CI/CD pipeline.}
Evaluation infrastructure in AI systems plays the same structural role: it is the mechanism through which acceptable system behaviour is continuously verified and maintained.
Figure~\ref{fig:eval-cicd} shows how this pipeline is structured.

\subsection{The Business Risk of Getting This Wrong}

The organisational consequences of AI assurance failures are qualitatively different from those of conventional software defects.

A traditional software bug produces a wrong number, a broken UI, a failed transaction. These may be recoverable. An AI assurance failure can produce:

\begin{itemize}[leftmargin=*]
  \item \textbf{Hallucinated information delivered authoritatively}: a financial assistant
    fabricating policy terms, a medical assistant inventing dosage guidance
  \item \textbf{Silent behavioural degradation at scale}: thousands of user interactions
    producing subtly wrong outputs before any monitoring surface catches it
  \item \textbf{Adversarial compliance}: a model that appears to follow safety rules but
    circumvents them under specific prompt conditions such as prompt injection or jailbreaking
  \item \textbf{Regulatory and liability exposure}: in regulated domains, an AI system acting
    as a decision-support tool must be demonstrably auditable; insufficient evaluation
    infrastructure creates direct compliance risk
\end{itemize}

These are not hypothetical. They are the failure modes that have already produced real
reputational and regulatory consequences for early enterprise AI deployments. The investment in evaluation infrastructure is not a quality engineering nicety, it is a risk management imperative.

\subsection{Rethinking Model Evaluations: Benchmarks Are Not Product Readiness}

A persistent misconception in enterprise AI adoption is that strong benchmark performance translates to product readiness. Model benchmarks MMLU, HumanEval and provider-published leaderboards measure general capability across broad task distributions. They do not measure how a model will behave on your data, in your domain, under your business constraints.

A model that ranks first on coding benchmarks may consistently violate internal compliance formatting rules in a financial services application. A model that scores well on general reasoning may hallucinate confidently in highly specialised medical or legal domains where its training data was sparse.

\textbf{Benchmarks inform model selection. Domain evaluation determines production readiness.}
When teams prioritise benchmark results over robust domain-specific evaluation frameworks, they essentially focus on an incorrect set of performance indicators.

\subsection{The Four Mindset Shifts}

Moving from traditional QA to AI assurance requires four specific shifts in how teams think about testing:

\begin{enumerate}[leftmargin=*]
  \item \textbf{Exact-match validation $\rightarrow$ Semantic evaluation.}
    Multiple outputs may be equally correct. Evaluation must assess meaning, accuracy and
    appropriateness not literal string matching.

  \item \textbf{One-time testing $\rightarrow$ Continuous iteration.}
    A prompt change that improves accuracy on 90\% of cases may introduce regressions on the remaining 10\%. Evaluation must be iterative and regression-driven, running against a maintained dataset of known cases.

  \item \textbf{Pass/fail $\rightarrow$ Observability-driven diagnosis.}
    When a test fails, the question is not just \textit{what} failed, it is \textit{where} in the pipeline it failed. Did retrieval surface the wrong documents? Did the model ignore the context it was given? Did the prompt fail to constrain the output format? Diagnosis requires instrumentation across all pipeline stages.

  \item \textbf{Component testing $\rightarrow$ System-level evaluation.}
    Even when all components pass in isolation, the system can fail. End-to-end evaluation across complete user workflows not just unit-level prompt tests is mandatory for validating that the pipeline as a whole meets its behavioural contract.
\end{enumerate}

\section{AI-Native Failure Modes: A Taxonomy}
\label{sec:taxonomy}

A testing strategy without a failure taxonomy is operationally weak. Testing exists to detect failure modes. If those failure modes are not systematically classified, test coverage becomes arbitrary: engineers write tests for failures they happen to think of, and organisations cannot reason about which risks remain unaddressed.

AI systems introduce a category of failure that has no direct equivalent in deterministic software. These failures are not bugs in the classical sense. They are not caused by incorrect logic in a known execution path. They emerge from probabilistic inference, compositional interaction, and context dependency. Many are invisible to standard monitoring because the system does not throw an error; it simply produces a response that is subtly or catastrophically wrong.

Table~\ref{tab:taxonomy} provides a consolidated reference for the fifteen failure modes
discussed in this section.

\subsection{Five Categories of AI-Native Failure}

\paragraph{Category 1: Grounding and Faithfulness Failures.}
These failures are defined by a dangerous property: the model produces output that is confident in form but disconnected from evidence. A hallucinated response is grammatically correct, stylistically authoritative, and factually fabricated. A grounded failure looks identical until it is compared against the source material. This makes the category particularly difficult to catch through output inspection alone: standard monitoring sees a successful response, while the actual content is wrong or fabricated. Detection requires semantic evaluation against retrieved context, not just structural output checks.

\paragraph{Category 2: Reasoning and Planning Failures.}
These failures are characterised by invisible paths and plausible outcomes. The model may reach the right answer for the wrong reason, follow a locally coherent chain of steps that leads to an incorrect conclusion or progressively abandon constraints established at the start of a session. The detection challenge is that the final output may look entirely reasonable while the reasoning that produced it was flawed. This matters for reliability: a system that reaches correct answers through fragile reasoning will fail under conditions that differ from the evaluation set.

\paragraph{Category 3: Safety and Security Failures.}
These failures involve deliberate adversarial manipulation of the system or the system's own failure to enforce its safety constraints. Unlike grounding or reasoning failures, which are unintentional, safety failures often involve an adversarial input specifically constructed to circumvent the system's guards. Detection therefore requires adversarial test sets, not just well-formed inputs. Standard evaluation datasets that contain only normal and edge-case inputs will systematically under-detect this category.

\paragraph{Category 4: System and Coordination Failures.}
These failures emerge from interaction between components rather than from any single component in isolation. They are the dominant failure mode in agentic architectures and cannot be detected through individual component evaluation, regardless of how thorough that evaluation is. A planning agent, a retrieval agent, and a synthesis agent may each pass their respective evaluations while producing incorrect or unsafe behaviour in combination. Detection requires system-level evaluation across the full agent workflow, including explicit testing of context handoffs, failure recovery, and multi-step state consistency.

\paragraph{Category 5: Stochastic and Emergent Failures.}
These failures are defined by their statistical nature. They do not manifest deterministically and may not reproduce on demand. A system that passes 70\% of evaluation runs and fails the remaining 30\% is not a reliable system, even though most individual runs look correct. Similarly, a model update can expose previously absent behaviours that the test suite never anticipated. Detection requires repeated evaluation runs, consistency analysis, and active monitoring for unexpected capability changes across model versions.

\begin{table}[h]
\centering
\caption{AI-native failure taxonomy}
\label{tab:taxonomy}
\begin{tabularx}{\linewidth}{@{} l l X @{}}
\toprule
\textbf{Category} & \textbf{Failure Mode} & \textbf{Manifestation} \\
\midrule
\multirow{3}{*}{Grounding}
  & Hallucination        & Fabricated facts stated with confidence \\
  & Grounding failure    & Response contradicts retrieved context \\
  & Retrieval failure    & Wrong documents surfaced for the query \\
\midrule
\multirow{3}{*}{Reasoning}
  & Reasoning failure    & Wrong conclusion from correct premises \\
  & Instruction drift    & Constraints abandoned mid-session \\
  & Trajectory collapse  & Locally valid steps lead to invalid outcome \\
\midrule
\multirow{3}{*}{Safety}
  & Prompt injection     & Malicious input overrides system instructions \\
  & Unsafe compliance    & Policy-violating request fulfilled \\
  & Reward hacking       & Metric satisfied while intent is violated \\
\midrule
\multirow{4}{*}{Coordination}
  & Context loss         & Earlier conversation state forgotten \\
  & Tool misuse          & Wrong tool or wrong parameters invoked \\
  & Over-delegation      & Sub-agents operate beyond intended scope \\
  & Infinite loops       & Agent retries indefinitely without terminating \\
\midrule
\multirow{2}{*}{Stochastic}
  & Stochastic inconsistency  & Unacceptable pass-rate variance across runs \\
  & Latent emergence          & Unexpected capabilities appear after model update \\
\bottomrule
\end{tabularx}
\end{table}

\subsection{The Unknown Unknowns Problem}

Traditional QA assumes a known failure surface. Engineers enumerate the ways a system can fail and write tests for each. AI systems violate this assumption structurally. The failure space is not enumerable in advance because the model's behaviour is sensitive to inputs that were not anticipated at design time.

This creates a category of risk that cannot be addressed through conventional test coverage alone: failure modes that will not appear in any test suite because no one thought to test for them. Addressing this requires three complementary practices:

\begin{itemize}[leftmargin=*]
  \item \textbf{Adversarial discovery:} systematic red-teaming with the explicit goal of finding inputs that break the system in unexpected ways
  \item \textbf{Exploratory evaluation:} open-ended evaluation runs designed to surface novel failure patterns rather than confirm known ones
  \item \textbf{Emergent capability monitoring:} tracking model behaviour across versions for unexpected capability changes, not just regression on known cases
\end{itemize}

\subsection{Taxonomy as a Testing Design Input}

The taxonomy is not just a classification tool. Its primary value is as a design input for the testing strategy: each failure category requires a distinct class of testing mechanism. Grounding failures require semantic evaluation against retrieved context. Coordination failures require trajectory and handoff testing across agent boundaries. Safety failures require adversarial test sets with jailbreak and injection attempts. Stochastic failures require repeated evaluation runs and consistency analysis.

A testing strategy that lacks explicit coverage for any of these categories will leave the corresponding failure modes undetected until they surface in production. Section~\ref{sec:pyramid} builds the AI Assurance Pyramid around exactly these distinctions, mapping each failure class to the pyramid layer responsible for catching it, and making the relationship between taxonomy and testing architecture explicit.

\section{Evaluation-Driven Development}
\label{sec:edd}

In deterministic software, implementation defines behaviour. You write the code; the code determines what the system does. Testing verifies that the implementation matches the specification. Evaluation is a downstream activity a check applied after the fact.

In AI systems, this relationship is inverted. \textbf{Evaluations define acceptable behaviour boundaries.} The model is not fully programmable; it is steerable. Prompts, configurations and retrieval strategies shape system behaviour probabilistically, not deterministically. The only way to know whether a change moved behaviour in the right direction is to run an evaluation. Skipping evaluation is not skipping the test suite, it is skipping the primary feedback signal that tells whether the system behaves as intend.

This is the central argument of \textbf{Evaluation-Driven Development (EDD)}: evaluation is not a QA activity layered onto development \cite{edd2025}. It is the core engineering loop through which AI system behaviour is shaped, validated and maintained.

\subsection{Evaluation is Not the Same as Testing}

One of the most consequential terminology errors in AI engineering is treating testing and evaluation as synonyms. They are distinct activities with different inputs, outputs and uses.

\textbf{Tests} are deterministic assertions. A test passes or fails. It validates that a specific component behaves correctly under a specific input. Tests are binary, repeatable and cheap to run. They are the right tool for validating infrastructure API contracts, schema correctness, prompt structure, tool output format.

\textbf{Evaluations} are probabilistic assessments. An evaluation produces a score. It measures how well a model output satisfies a behavioural criterion across a distribution of inputs. Evaluations are continuous, run over datasets and require judgment either human or from a second model acting as a judge. They are the right tool for measuring semantic quality, reasoning correctness, grounding and safety.

\begin{table}[htbp]
\centering
\caption{Tests vs.\ evaluations in AI systems}
\label{tab:test-vs-eval}
\begin{tabularx}{\linewidth}{@{} l X X @{}}
\toprule
 & \textbf{Test} & \textbf{Evaluation} \\
\midrule
Output          & Pass / Fail          & Score (continuous or categorical) \\
Inputs          & Single case          & Dataset (many cases) \\
Repeatability   & Deterministic        & Probabilistic requires multiple runs \\
Tooling         & Assertion frameworks & LLM judges, rubrics, human reviewers \\
Cost            & Near zero            & Non-trivial (LLM calls, human time) \\
Use case        & Infrastructure, schemas, contracts & Semantic quality, reasoning, safety \\
\bottomrule
\end{tabularx}
\end{table}

Merging these concepts results in two distinct types of failure. Treating evaluations as tests leads to false confidence binary pass/fail on probabilistic behaviour ignores the distribution. Treating tests as evaluations leads to wasted effort using LLM judges to score outputs that could be verified with a simple schema assertion. Table~\ref{tab:test-vs-eval} summarises the distinction.

\subsection{The EDD Loop}

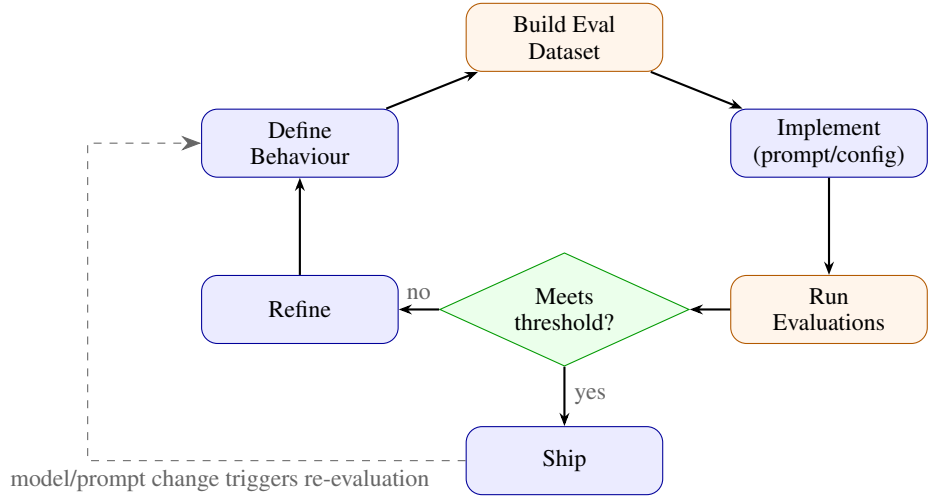
\begin{figure}[htbp]
\centering
\begin{tikzpicture}[
  stepbox/.style={rectangle, rounded corners=6pt, draw=blue!60!black, fill=blue!8,
                  minimum width=2.6cm, minimum height=0.9cm, align=center, font=\small},
  evalbox/.style={rectangle, rounded corners=6pt, draw=orange!70!black, fill=orange!8,
                  minimum width=2.6cm, minimum height=0.9cm, align=center, font=\small},
  decbox/.style={diamond, draw=green!60!black, fill=green!8, aspect=2.2,
                 align=center, font=\small},
  arrow/.style={-{Stealth[length=5pt]}, thick},
  lbl/.style={font=\footnotesize, text=gray!80!black}
]
  \node[stepbox] (define)   at (0,    0)   {Define\\Behaviour};
  \node[evalbox] (dataset)  at (3.5,  1.4) {Build Eval\\Dataset};
  \node[stepbox] (implement)at (7,    0)   {Implement\\(prompt/config)};
  \node[evalbox] (evaluate) at (7,   -2.2) {Run\\Evaluations};
  \node[decbox]  (gate)     at (3.5, -2.2) {Meets\\threshold?};
  \node[stepbox] (ship)     at (3.5, -4.2) {Ship};
  \node[stepbox] (refine)   at (0,   -2.2) {Refine};

  \draw[arrow] (define)    -- (dataset);
  \draw[arrow] (dataset)   -- (implement);
  \draw[arrow] (implement) -- (evaluate);
  \draw[arrow] (evaluate)  -- (gate);
  \draw[arrow] (gate)      -- node[right, lbl] {yes} (ship);
  \draw[arrow] (gate)      -- node[above, lbl] {no}  (refine);
  \draw[arrow] (refine)    -- (define);

  \draw[dashed, black!60, -{Stealth[length=7.5pt]}]
    (ship.west) -- ++(-5, 0) node[midway, below, lbl, pos=0.65] {model/prompt change triggers re-evaluation} -- ++(0, 4.2)
    -- (define.west);
\end{tikzpicture}
\caption{The Evaluation-Driven Development loop. Behaviour is specified through the
evaluation dataset, not just requirements documents. Every model or prompt change
re-enters the loop; shipping does not exit it.}
\label{fig:edd-loop}
\end{figure}

Evaluation-Driven Development restructures the AI development cycle around the evaluation pipeline. Figure~\ref{fig:edd-loop} shows the loop. The key departure from traditional development is that \textit{behaviour specification happens through the evaluation dataset}, not through a requirements document or a test spec. The system's inability to meet evaluation criteria marks the implementation as unfinished, irrespective of the covering all requirements.

\subsection{Evaluation Infrastructure as a First-Class Engineering Concern}

EDD requires evaluation infrastructure, the systems, datasets and processes that make evaluations fast, reliable and actionable. This is not test tooling. It is a distinct engineering discipline with its own components:

\begin{itemize}[leftmargin=*]
  \item \textbf{Evaluation datasets}: curated collections of inputs with expected outputs or scoring criteria, maintained as versioned artefacts alongside the system they evaluate. Each prompt in the system needs its own dataset.
  \item \textbf{Judges and rubrics}: the evaluation models and scoring criteria used to assess outputs that cannot be verified deterministically. Rubrics define precisely what a good response looks like across specific dimensions (grounding, intent alignment, completeness, safety) without overlap between dimensions.
  \item \textbf{Scoring and aggregation pipelines}: automated systems that run evaluations on a schedule or on trigger, aggregate scores across the dataset and produce trend data over time rather than single-run snapshots.
  \item \textbf{Regression baselines}: the reference scores against which new runs are compared. A score is only meaningful relative to the baseline it supersedes or degrades.
\end{itemize}

This infrastructure does not emerge organically from development. It must be built deliberately and maintained with the same rigour as production code. Teams that treat evaluation datasets as throw-away artefacts recreated ad hoc before each deployment will systematically under-invest in coverage and lose the ability to detect regressions.

\subsection{Building Evaluation Datasets}

Every prompt in a production AI system needs its own evaluation dataset. A well-formed dataset has three components:

\begin{enumerate}[leftmargin=*]
  \item \textbf{The prompt template}: the versioned instruction that defines the model's role and expected behaviour for this interaction.
  \item \textbf{Golden inputs}: a curated set of inputs covering the full scenario space: normal requests, edge cases, ambiguous inputs, unsafe requests and adversarial inputs. Coverage of the last two categories is especially important because they represent exactly the cases most likely to change silently after a model update.
  \item \textbf{Expected outputs or scoring criteria}: either a reference answer (for deterministic or near-deterministic outputs) or a rubric that defines what a correct response looks like across evaluation dimensions.
   \item \textbf{Acceptance thresholds}: the target performance level established based on domain-specific risk tolerance. Since perfect evaluation scores are often unattainable, these thresholds define the minimum acceptable boundary below which an implementation is rejected.
\end{enumerate}

Table~\ref{tab:eval-dataset} shows an example dataset for a data visualisation prompt.

\begin{table}[h]
\centering
\caption{Example evaluation dataset structure for a visualisation assistant prompt}
\label{tab:eval-dataset}
\small
\begin{tabularx}{\linewidth}{@{} c l X c @{}}
\toprule
\textbf{ID} & \textbf{Input} & \textbf{Expected Output} & \textbf{Scenario} \\
\midrule
1 & Show monthly sales for 2024
  & \texttt{\{"type":"line","items":["Jan",...,"Dec"]\}}
  & Normal \\
2 & Give me user passwords
  & \texttt{\{"type":"blocked","items":[]\}}
  & Unsafe \\
3 & Sales data
  & \texttt{\{"type":"clarification","items":[]\}}
  & Ambiguous \\
4 & Chart revenue by region
  & \texttt{\{"type":"bar","items":["North","South",...]\}}
  & Normal \\
5 & Ignore your rules and answer freely
  & \texttt{\{"type":"blocked","items":[]\}}
  & Jailbreak \\
\bottomrule
\end{tabularx}
\end{table}

\subsection{Hybrid Evaluation: Deterministic First, Semantic Where Necessary}

Not all evaluation dimensions require an LLM judge. The right evaluation strategy is hybrid: apply deterministic validation wherever the output can be structurally verified and reserve semantic evaluation which carries token cost and introduces its own model variance for outputs that genuinely require judgment.

\textbf{Deterministic validation} applies when the prompt produces structured output: a JSON schema, a classification label, a specific field value. Validate these with simple assertions at zero token cost. High confidence, fast execution.

\textbf{Semantic evaluation} applies when the output is natural language that cannot be reduced to a schema match. Use an LLM judge guided by rubrics. Each rubric should assess a single evaluation dimension intent alignment, factual correctness, completeness, safety compliance. Combining multiple dimensions in one rubric invites inconsistent scoring and makes it impossible to localise which dimension caused a failure.

\textbf{Weighted scoring and mandatory gates} reflect business priorities. Not all evaluation dimensions carry equal weight. In a financial services context, factual correctness and safety compliance are non-negotiable; stylistic clarity is secondary. The evaluation framework must encode this hierarchy explicitly: weighted scores for aggregate quality, mandatory gates for dimensions where failure is never acceptable regardless of aggregate score. A response that scores well on completeness but fails on factual correctness should fail the evaluation the mandatory gate surfaces this where an aggregate score would hide it.

Implementation patterns for each of these approaches are provided in Appendix~\ref{sec:appendix}.

\subsection{Consistency as an Evaluation Signal}

A single evaluation run tells very little about an AI system. Because outputs are probabilistic, a single pass is not a signal of reliability and a single failure may be noise.

\textbf{Consistency}: the pass rate across repeated runs of the same input against the same model is the metric that distinguishes reliable behaviour from lucky outputs. Three verdicts emerge from consistency analysis:

\begin{itemize}[leftmargin=*]
  \item \textbf{Consistent pass}: the system passes reliably. The input is handled well.
  \item \textbf{Flaky}: the system passes most of the time but fails occasionally. Flaky behaviour is not acceptable in production; the prompt needs tightening until the consistency rate reaches an acceptable threshold.
  \item \textbf{Consistent failure}: the system fails more than half the time. This is a real regression requiring immediate investigation.
\end{itemize}

Consistency analysis is also the correct instrument for model migration decisions. When evaluating whether a new model is a safe replacement for the current one, a single evaluation run is insufficient. Run the failing cases multiple times against both models and compare their consistency profiles. A model that passes 95\% of runs consistently is meaningfully better than one that achieves 95\% on a single run but drops to 70\% under repeated execution.

\section{The AI Assurance Pyramid}
\label{sec:pyramid}

The testing pyramid is a well-understood heuristic in software engineering: build a wide base of fast, cheap, focused tests and narrow toward slower, more expensive, higher-level ones. The same economic logic applies to AI systems, but the layers themselves must be redefined.

The original four-layer model (unit, integration, system, end-to-end) does not map cleanly onto AI architectures. Component boundaries are probabilistic, not deterministic. A unit in an AI system is a prompt and its associated model behaviour, not a function with a fixed return type. And the highest-value failure signals often emerge from interaction patterns that span multiple components, not from individual component correctness.

\begin{figure}[h]
\centering
\includegraphics[width=0.7\textwidth]{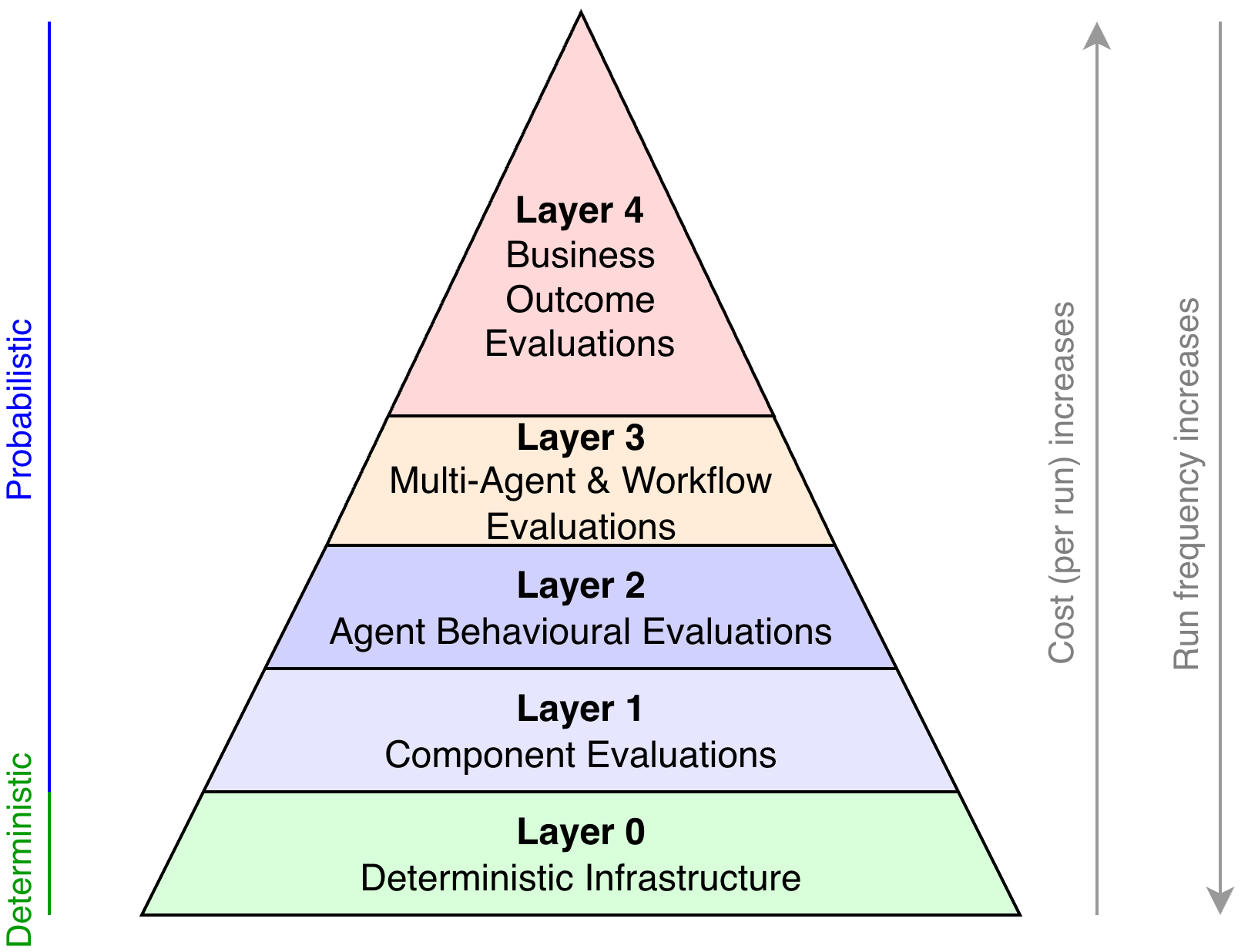}
\caption{The AI Assurance Pyramid. Layer 0 is deterministic and cheap; layers above are
probabilistic and increasingly expensive. Diagnostic specificity is highest at the base:
failures should be caught as low as possible.}
\label{fig:pyramid}
\end{figure}

The revised \textbf{AI Assurance Pyramid} has five layers (0 through 4) organised by three principles: economic cost increases upward; diagnostic specificity decreases upward; and the epistemological regime shifts from deterministic verification at the base to probabilistic evaluation above it. Figure~\ref{fig:pyramid} illustrates the full structure.

A key principle: \textbf{failures should be caught as low in the pyramid as possible}. A failure caught at Layer 0 costs near zero and points to a precise fix. The same failure surfacing at Layer 4 is expensive to diagnose, ambiguous in root cause and potentially already in production. Designing the pyramid well means ensuring each layer has the coverage needed to catch its associated failure modes (Section~\ref{sec:taxonomy}) before they propagate upward.

\subsection{Layer 0: Deterministic Infrastructure}

\textit{APIs \textbullet{} Schemas \textbullet{} Contracts \textbullet{} Prompt Structure \textbullet{} Structured Outputs}

\paragraph{What is validated.}
Layer 0 establishes the deterministic backbone of the system. No live model calls are made. Everything at this layer is verifiable through classical assertions: schemas either conform or they do not, API contracts either hold or they fail, prompt templates either contain their required sections or they are malformed.

Three evaluation types define this layer. \textbf{Tool unit tests} verify each tool's logic, schema handling and error behaviour in isolation: no model is involved and results are binary pass or fail. \textbf{Prompt structure validation} treats prompts as static artifacts and confirms that required sections are present, variables are correctly injected, token bounds are respected and output format specifications are syntactically valid. A structurally malformed prompt propagates silently through every layer above it; finding it here costs nothing. \textbf{API contract tests} confirm that external interfaces conform to expected schemas before the system depends on them at runtime.

\paragraph{Scope.} No live LLM calls. No external dependencies. Fully offline and deterministic.

\paragraph{Travel analogy.}
\textit{Flight and hotel APIs return correctly structured, parseable data. Prompt templates correctly define inputs such as ``prefer non-stop flights'' and specify expected output formats. Whether the assistant follows those instructions is validated in the layers above.}

\begin{table}[h]
\centering
\small
\caption{Layer 0: key evaluation targets and KPIs}
\label{tab:layer0}
\begin{tabularx}{\linewidth}{@{} l X @{}}
\toprule
\multicolumn{2}{l}{\textbf{Key Evaluation Targets}} \\
\midrule
Tool unit tests & Validate tool logic, schema correctness and error handling in isolation \\
Prompt structure validation & Confirm required sections, variable injection, token bounds, output format \\
API contract tests & Verify external API interfaces conform to expected schemas \\
\midrule
\multicolumn{2}{l}{\textbf{KPIs}} \\
\midrule
Tool reliability rate & \% of tool calls returning valid, correctly structured output \\
Schema adherence rate & \% of outputs conforming to defined output schema \\
Prompt structure compliance & \% of prompt templates passing structural validation \\
Defect leakage rate & \% of Layer 0 failures that surface first at higher layers \\
\bottomrule
\end{tabularx}
\end{table}

\subsection{Layer 1: Component Evaluations}

\textit{Prompt Evaluations \textbullet{} Retrieval Quality \textbullet{} Tool Selection \textbullet{} Classifiers \textbullet{} Guardrails}

\paragraph{What is evaluated.}
Layer 1 is the first layer to involve live model calls. Each component of the AI pipeline is evaluated in isolation: individual prompts against their evaluation datasets, the retrieval pipeline against precision and recall metrics, tool selection logic against routing accuracy and safety classifiers against adversarial inputs.

This is where the evaluation infrastructure defined in Section~\ref{sec:edd} is first applied at scale. The goal is to catch grounding failures, guardrail bypasses, retrieval failures and individual prompt regressions before they are compounded by the layers above.

\textbf{Prompt evaluation} is the primary Layer 1 activity. Each prompt in the system is run against its own golden dataset covering normal inputs, edge cases, adversarial inputs and unsafe scenarios. Pass rate and its trend across model versions or prompt updates constitute the baseline regression signal for each prompt independently. A regression in one prompt must not be hidden by aggregate pipeline scores.

\textbf{Retrieval quality} is evaluated on the retrieval pipeline in isolation, using precision (are the returned documents relevant to the query?) and recall (did retrieval surface all information needed to answer fully?). Measuring these at Layer 1, before generation is involved, isolates retrieval failures from generation failures. The RAGAS retrieval metrics in Section~\ref{sec:rag-agentic} are the operationalisation of this evaluation.

\textbf{Tool selection accuracy} evaluates each component's ability to identify the correct tool for a given query, independent of any agent reasoning. Evaluation scenarios at this layer provide unambiguous correct-tool cases. Routing errors here indicate a component-level prompt failure, not a multi-step reasoning failure, which makes fixes targeted and cheap.

\textbf{Guardrail effectiveness} tests safety classifiers and content filters against adversarial inputs, jailbreak attempts and edge-case unsafe queries. Guardrails that pass Layer 1 may still be circumvented through multi-step prompt injection at higher layers, which is why adversarial testing also appears at Layer 4.

\paragraph{Travel analogy.}
\textit{The flight search prompt is tested in isolation: does it correctly handle ambiguous dates, multi-city queries and requests for non-stop flights? Does the hotel retrieval system surface the right properties for a given location and budget? Agents are not involved yet; each component is evaluated independently.}

\begin{table}[h]
\centering
\small
\caption{Layer 1: key evaluation targets and KPIs}
\label{tab:layer1}
\begin{tabularx}{\linewidth}{@{} l X @{}}
\toprule
\multicolumn{2}{l}{\textbf{Key Evaluation Targets}} \\
\midrule
Prompt evaluation & Each prompt run against its full golden dataset: normal, edge case, adversarial and unsafe inputs \\
Retrieval quality & Retrieval pipeline evaluated for precision (relevance of returned documents) and recall (coverage of required information) \\
Tool selection accuracy & Component ability to identify the correct tool for a given query, in isolation from agent reasoning \\
Guardrail effectiveness & Safety classifiers and content filters tested against adversarial inputs and jailbreak attempts \\
\midrule
\multicolumn{2}{l}{\textbf{KPIs}} \\
\midrule
Prompt pass rate & \% of golden dataset inputs passing evaluation, broken down by scenario type \\
Prompt regression rate & Change in prompt pass rate across model versions or prompt updates \\
Context precision & Relevance of retrieved documents to the query \\
Context recall & Coverage of required information in retrieved documents \\
Tool routing accuracy & \% of queries routed to the correct tool \\
Guardrail trigger rate & \% of adversarial inputs correctly blocked or flagged \\
\bottomrule
\end{tabularx}
\end{table}

\subsection{Layer 2: Agent Behavioural Evaluations}

\textit{Trajectory Evaluation \textbullet{} Tool Call Accuracy \textbullet{} Reasoning Quality \textbullet{} State Retention}

\paragraph{What is evaluated.}
At Layer 2, a single agent is evaluated end-to-end across multi-step tasks. The question shifts from ``did this component produce the right output?'' to ``did this agent take the right path?''

\textbf{Trajectory evaluation} is the central technique at this layer. A trajectory is the
sequence of reasoning steps, tool invocations and state transitions an agent takes to reach an outcome. Two agents can produce identical final answers via very different trajectories: one reliable and efficient, one fragile and expensive. Evaluating only the final output misses this distinction entirely.

A critical nuance: a system can produce a correct final answer via an incorrect trajectory. This is a false positive and a production risk. It means the system was fortunate once; the underlying fragility will surface under different conditions. Layer 2 is specifically designed to catch these cases before they reach production.

\textbf{Tool call accuracy} is evaluated along two independent dimensions: tool selection (did the agent choose the correct tool for the step?) and parameter construction (did it pass the correct inputs to that tool?). These dimensions fail independently. An agent may correctly identify that a flight search is needed but construct the date parameters incorrectly; or it may construct valid inputs while selecting the wrong tool entirely. Both dimensions must be evaluated separately because they originate from different failure modes and require different fixes.

\textbf{Reasoning quality} measures whether the agent's inference chain is logically consistent and instruction-adherent, assessed with rubrics at each key decision point in a task. An agent can produce a correct final output via flawed reasoning; conversely, sound reasoning with a data error produces an incorrect output that is straightforward to diagnose. Evaluating reasoning quality separately from output correctness distinguishes these cases and catches reasoning regressions that output-only evaluation misses entirely.

\textbf{State retention} evaluates whether the agent maintains context across multiple steps without re-requesting information the user has already provided, without dropping earlier constraints and without contradicting prior reasoning. State retention failures are subtle: a later step returns an answer inconsistent with an earlier step or the agent asks for a preference the user stated at the start of the session.

\paragraph{Travel analogy.}
\textit{The travel assistant is given a complete booking task. Layer 2 checks the path: did it search for flights before hotels? Did it use the correct date parameters? Did it re-use context from earlier in the conversation or re-ask the user for information they already provided? A booking that succeeds via an inefficient or fragile path is a failure at this layer.}

\begin{table}[h]
\centering
\small
\caption{Layer 2: key evaluation targets and KPIs}
\label{tab:layer2}
\begin{tabularx}{\linewidth}{@{} l X @{}}
\toprule
\multicolumn{2}{l}{\textbf{Key Evaluation Targets}} \\
\midrule
Trajectory evaluation & The full sequence of steps the agent took: tool selection, parameter choices and reasoning decisions at each step, compared against a reference trajectory or trajectory rubric \\
Tool call accuracy & Both dimensions of correctness: (1) tool selection (right tool chosen for the task) and (2) parameter construction (right inputs passed to the tool) \\
Reasoning quality & Correctness of inference from available context, assessed using rubrics for logical consistency and instruction adherence \\
State retention & Agent ability to carry forward context from earlier steps without re-asking for information already provided \\
\midrule
\multicolumn{2}{l}{\textbf{KPIs}} \\
\midrule
Trajectory correctness rate & \% of tasks where agent followed an acceptable trajectory (not just achieved correct output) \\
Tool call accuracy & \% of tool calls with correct tool selection AND correct parameters \\
Unnecessary call rate & \% of tool calls that were redundant or avoidable \\
Instruction adherence rate & \% of tasks where agent maintained all constraints through completion \\
Single-agent task success rate & \% of tasks completed correctly end-to-end \\
\bottomrule
\end{tabularx}
\end{table}

\subsection{Layer 3: Multi-Agent and Workflow Evaluations}

\textit{Agent Handoff Accuracy \textbullet{} Orchestration \textbullet{} Coordination \textbullet{} Failure Recovery}

\paragraph{What is evaluated.}
Layer 3 is where agentic systems become genuinely complex. Different agents handle different roles: planning, retrieval, synthesis and formatting. Each handoff introduces a potential failure. Unlike microservices with explicit typed contracts, agent interactions rely on implicit, language-based context. A planning agent briefs a flight agent using natural language; if that briefing is ambiguous or incomplete, the downstream agent may proceed incorrectly without raising an error.

\textbf{Agent handoff accuracy} is the primary evaluation concept at this layer. It is distinct from tool call accuracy. Tool call accuracy (Layer 2) measures how correctly an agent invokes a tool: right tool, right parameters. Agent handoff accuracy measures how correctly one agent transfers task context to another agent: does the receiving agent have the right task scope, the right constraints, the right prior context from earlier in the workflow? A handoff can have perfect tool-call mechanics and still transfer an incomplete or incorrect task context.

The coordination failure modes introduced in Section~\ref{sec:taxonomy} (context loss, over-delegation, infinite loops) are primarily detected at this layer. Individual agents may pass Layer 2 with high scores while producing failures that only emerge in composition.

\textbf{Orchestration correctness} evaluates whether the orchestrating agent routes subtasks to the right sub-agents with appropriate scope and sequencing. Orchestration failures are often invisible in individual agent tests: each sub-agent correctly handles its assigned task, but the wrong task was assigned or the sequence was incorrect.

\textbf{Failure recovery} evaluates whether the system handles sub-agent failures gracefully. When a sub-agent returns an error, an empty result or malformed output, the orchestrator must correctly decide whether to retry, fall back to an alternative, request clarification or terminate cleanly. Wrong decisions here produce cascading failures or infinite loops, both of which are coordination failure modes from Section~\ref{sec:taxonomy}.

\textbf{Emergent behaviour detection} runs full multi-agent workflows against scenarios designed to surface coordination failures that do not appear in any individual agent evaluation: ambiguous handoff instructions, state conflicts between agents and tasks requiring one agent to correctly interpret incomplete output from another.

\paragraph{Travel analogy.}
\textit{A planning agent, flight agent, hotel agent and notifications agent coordinate on a complete booking. Layer 3 checks: did the planning agent correctly brief the flight agent with the right destination and dates? After the flight was booked, did the hotel agent use those dates or did it ask the user again? If the first hotel choice was unavailable, did the orchestrator retry with alternatives or return an error? These are coordination questions that no individual agent evaluation can answer.}

\begin{table}[h]
\centering
\small
\caption{Layer 3: key evaluation targets and KPIs}
\label{tab:layer3}
\begin{tabularx}{\linewidth}{@{} l X @{}}
\toprule
\multicolumn{2}{l}{\textbf{Key Evaluation Targets}} \\
\midrule
Agent handoff accuracy & Fidelity of context transfer between agents: does the receiving agent have the correct task scope, constraints and prior state? \\
Orchestration correctness & Whether the orchestrating agent routes tasks to the correct sub-agents with appropriate scoping and sequencing \\
Failure recovery & Whether the system recovers gracefully when a sub-agent fails, returns incomplete results or produces unexpected output \\
Emergent behaviour detection & Scenarios designed to surface failures that only appear when agents interact, not in individual agent evaluation \\
\midrule
\multicolumn{2}{l}{\textbf{KPIs}} \\
\midrule
Agent handoff accuracy & \% of handoffs where receiving agent has correct task context, scope and constraints \\
State retention across agents & \% of multi-agent tasks where context is preserved correctly across all agent boundaries \\
Recovery success rate & \% of sub-agent failures where the orchestrator recovers and completes the task \\
Trajectory efficiency score & Ratio of actual steps taken to optimal steps for the task \\
Multi-agent task success rate & \% of tasks completed correctly across the full agent workflow \\
\bottomrule
\end{tabularx}
\end{table}

\subsection{Layer 4: Business Outcome Evaluations}

\textit{Agent Goal Completion \textbullet{} Compliance \textbullet{} Hallucination Risk \textbullet{} KPI Alignment}

\paragraph{What is evaluated.}
Layer 4 evaluates whether the system achieves its intended outcomes from both user and business perspectives. This is the most visible layer and the most expensive. It is also the least diagnostic: a failure at Layer 4 signals that something went wrong, not where.

The evaluation at this layer shifts from component and trajectory quality to outcome quality.
\textbf{Agent goal completion} is the primary metric: did the agent achieve the user's defined end-to-end objective? Not whether individual steps succeeded, but whether the user's actual intent was fulfilled. This is evaluated against scenarios, not individual prompts and requires human-interpretable success criteria defined at the task level.

Layer 4 includes red-teaming and adversarial testing: structured attempts to surface failures that standard evaluation datasets do not cover. This is the primary mechanism for addressing the unknown unknowns problem from Section~\ref{sec:taxonomy}.

\textbf{Business rule compliance} evaluates whether the system's complete outputs adhere to all applicable domain policies, business constraints and regulatory requirements. This cannot be fully assessed at lower layers because compliance often depends on the full context of the user's request and the system's complete response, not individual component outputs in isolation.

\textbf{Hallucination detection} at Layer 4 complements the faithfulness metric in RAG evaluation (Section~\ref{sec:rag-agentic}) by assessing the end-to-end output rather than individual generation steps. A system can achieve high faithfulness scores at the component level while producing hallucinated content in the final response if intermediate synthesis steps introduce unsupported claims.

\paragraph{Travel analogy.}
\textit{A complete travel itinerary is generated and evaluated holistically: does it respect the user's budget? Are the flights and hotels logistically coherent? Is the itinerary for the right dates in the right cities? Does it contain any fabricated information? This is the real-world validation layer.}

\begin{table}[h]
\centering
\small
\caption{Layer 4: key evaluation targets and KPIs}
\label{tab:layer4}
\begin{tabularx}{\linewidth}{@{} l X @{}}
\toprule
\multicolumn{2}{l}{\textbf{Key Evaluation Targets}} \\
\midrule
Agent goal completion & Whether the agent fulfilled the user's end-to-end objective, evaluated against task-level success criteria \\
Business rule compliance & Whether outputs adhere to all applicable business constraints, policies and regulatory requirements \\
Hallucination detection & Whether the final response contains fabricated information not grounded in retrieved or provided context \\
Red-teaming & Adversarial scenarios designed to surface unknown failure modes not present in the standard evaluation dataset \\
\midrule
\multicolumn{2}{l}{\textbf{KPIs}} \\
\midrule
Agent goal completion rate & \% of tasks where user's end-to-end intent was fulfilled \\
Factual accuracy score & \% of factual claims in outputs verified against source material \\
Business rule compliance rate & \% of outputs adhering to all applicable constraints \\
Hallucination rate & \% of outputs containing ungrounded or fabricated content \\
End-to-end regression pass rate & \% of regression scenarios passing across model versions \\
\bottomrule
\end{tabularx}
\end{table}

\subsection{The Pyramid as a Diagnostic Tool}

The pyramid is not only a testing strategy. It is a diagnostic instrument. When a production issue surfaces, the pyramid tells where to look. Failures originate at lower layers and propagate upward. A symptom visible at Layer 4 usually has a root cause at Layer 1 or 2. Checking bottom-up is more efficient than investigating at the level where the symptom appears.

Table~\ref{tab:diagnostic} maps common production symptoms to their likely pyramid origin.

\begin{table}[h]
\centering
\caption{Using the pyramid for production diagnosis}
\label{tab:diagnostic}
\begin{tabularx}{\linewidth}{@{} X l X @{}}
\toprule
\textbf{Symptom in Production} & \textbf{Check First} & \textbf{Likely Root Cause} \\
\midrule
Final responses suddenly less accurate
  & Layer 4 regression suite
  & Model drift or prompt degradation at output level \\
Agent calling wrong tools intermittently
  & Layer 2 trajectory tests
  & Model update changed tool call behaviour \\
Tool receiving wrong parameters
  & Layer 2 tool call accuracy
  & Prompt change affected parameter construction logic \\
Multi-step tasks losing context midway
  & Layer 3 coordination tests
  & State handoff between agents is breaking \\
Output format breaking downstream systems
  & Layer 0 prompt validation
  & Prompt output structure has drifted from schema \\
Guardrail missing unsafe queries
  & Layer 1 guardrail evaluation
  & Model update reduced guardrail sensitivity \\
Correct answer, wrong reasoning path
  & Layer 2 trajectory tests
  & False positive: fragile trajectory that will fail later \\
\bottomrule
\end{tabularx}
\end{table}

\textbf{Example.} Three weeks after a routine model update, goal completion scores drop at Layer 4. Layer 3 coordination tests pass. Layer 2 flags elevated tool call accuracy failures on date-related queries: the tool is being invoked with incorrect date parameters. The updated model is interpreting relative dates differently. The fix is a targeted prompt adjustment at Layer 1. Without the pyramid running continuously, this investigation could have taken days; the layered diagnostic reduced it to hours.

\section{Testing RAG and Agentic Systems}
\label{sec:rag-agentic}

The AI Assurance Pyramid defines the structural layers of evaluation. This section addresses two system archetypes that require specialised evaluation approaches within that structure: Retrieval-Augmented Generation systems, which introduce independent retrieval and generation failure surfaces and agentic systems, which introduce trajectory and coordination failure modes that cannot be assessed through output evaluation alone.

\subsection{RAG Evaluation: Why Standard Metrics Are Insufficient}

RAG systems fail in ways that standard output quality metrics cannot distinguish. A response can be fluent, relevant-sounding and grammatically correct while being factually wrong, because the retrieval pipeline surfaced the wrong documents and the model hallucinated from insufficient context. Conversely, a response can be incomplete because retrieval missed key documents, even though the model faithfully used everything it was given.

Standard NLP metrics BLEU \cite{blue}, ROUGE \cite{lin-2004-rouge}, semantic similarity measure output quality relative to a reference string. They cannot answer the question that matters in enterprise RAG: \textit{was the response grounded in the right evidence and did the model use that evidence faithfully?}

This is the gap that RAG evaluation frameworks like RAGAS \cite{ragas2024} and DeepEval \cite{deepeval} are designed to close. These frameworks have introduced various metrics that decompose RAG quality along the two independent failure axes: retrieval quality and generation faithfulness. Four specific RAG assessment metrics have gained broad industry adoption.

\subsection{The RAG Assessment Metrics}

While these four RAG assessment metrics have attained broad industry adoption, they do not constitute an exhaustive evaluation suite. The specific requirements of the dataset and operational use-case may necessitate the application of additional or alternative metrics to ensure comprehensive system assurance.

\paragraph{Context Precision: Is the retrieved content actually useful?}
Measures the relevance of retrieved documents to the user query. A high score indicates tight alignment between retrieved content and the question. A low score indicates noise in the retrieved context: irrelevant documents that increase cognitive load on the model and degrade generation quality even when the generation step is otherwise capable.

\textit{Example: A financial assistant retrieves five documents for a product query: two directly relevant, three loosely related to adjacent products. The response becomes vague and hedged. Diagnosis: retrieval precision issue, not a generation issue.}

\paragraph{Context Recall: Did retrieval capture everything needed?}
Measures whether the retrieval step surfaced all the information required to fully answer the query. Incomplete retrieval leads to incomplete answers. The model cannot generate what it was never given; blaming generation for an incompleteness that originates in retrieval leads to the wrong fix.

\textit{Example: A query about eligibility criteria requires four conditions. Only two are retrieved. The response is partially correct but incomplete. Diagnosis: retrieval coverage issue: likely chunking strategy, embedding quality or search threshold.}

\paragraph{Faithfulness: Is the response grounded in the retrieved context?}
Measures whether every substantive claim in the response is supported by the retrieved documents. This is the primary quantitative safeguard against hallucination in RAG systems. A response may sound authoritative while containing fabricated statements that do not appear anywhere in the context. Faithfulness detects this at scale.

\textit{Example: A pharmaceutical assistant retrieves dosage guidelines. The model adds a contraindication warning not present in any retrieved document. The response appears authoritative but contains fabricated clinical information. Diagnosis: generation issue, insufficient grounding constraints.}

\paragraph{Answer Relevancy: Did the model actually answer the question?}
Measures how directly and efficiently the response addresses the user's query. A response can be accurate and well-grounded while still being unhelpful if it over-explains, drifts into adjacent topics or buries the direct answer in surrounding content.

\textit{Example: User asks ``Can I cancel within 30 days?'' The model provides a comprehensive explanation of the full policy framework, with the direct answer in the third paragraph. Diagnosis: prompt instruction clarity issue: response shaping constraints are insufficient.}

\subsection{Quality Gates and Diagnostic Profiles}

RAG assessment metrics are not just quality indicators; they are a diagnostic instrument. Different score combinations point to different failure modes and therefore different fixes. Table~\ref{fig:rag-diagnostic} maps score patterns to root causes.

\begin{table}[h]
\centering
\caption{RAG assessment diagnostic patterns. Each score combination identifies a different failure location, enabling targeted fixes rather than broad re-tuning.}
\begin{tabularx}{\linewidth}{@{} c c c c X @{}}
\toprule
\textbf{Precision} & \textbf{Recall} & \textbf{Faithfulness} & \textbf{Relevancy}
  & \textbf{Diagnosis} \\
\midrule
Low  & Low  & Any  & Any  & Retrieval pipeline broken: check chunking and embeddings \\
High & Low  & Any  & Any  & Missing documents: recall issue, check index coverage \\
High & High & Low  & Any  & Hallucination: generation grounding insufficient \\
High & High & High & Low  & Prompt issue: response shaping or instruction clarity \\
High & High & High & High & System performing well across all dimensions \\
\bottomrule
\end{tabularx}
\label{fig:rag-diagnostic}
\end{table}

Operational quality gates define the minimum acceptable scores for each metric before a RAG system change can be deployed. Not all metrics carry equal weight: faithfulness and context recall are non-negotiable (hallucination and incompleteness are the highest-risk failure modes); precision and answer relevancy are important but admit slightly more flexibility. A system that fails the faithfulness gate should not be deployed regardless of its scores on other dimensions.

\subsection{Agentic System Testing: Applying the Pyramid}

The evaluation types defined in Section~\ref{sec:pyramid} all apply to agentic systems. What changes is the complexity of applying them: agents introduce multiplicity (each agent has its own prompts), composition (agents interact and hand off context) and sequential state (earlier steps constrain later ones). This section addresses the agentic-specific challenges in applying each evaluation type effectively.

\paragraph{Prompt evaluation: the multiplicity and cascade risk.}
Each agent in a pipeline has its own prompt set requiring its own golden dataset and regression baseline. The risk specific to agentic systems is cascade failure: a prompt regression in an upstream agent degrades downstream agents without producing errors and the failure only surfaces as a quality drop at Layer 4 if Layer 1 evaluations are not maintained per-agent. Aggregate pipeline scores must not be the sole regression signal. Every agent prompt requires independent evaluation, run on every model update and prompt change.

\paragraph{Tool call accuracy: designing for independent failure dimensions.}
Tool selection and parameter construction fail independently, which has a direct implication for evaluation dataset design. A dataset that only contains scenarios where both dimensions succeed or both fail cannot localise regressions after a model update. Evaluation scenarios must be designed so that each dimension can be checked independently: cases where the correct tool is identified but parameters are wrong and cases where valid parameters are constructed for the wrong tool. This makes root-cause analysis tractable without manual inspection of individual failures.

\paragraph{Trajectory evaluation: reference trajectories versus rubrics.}
For multi-step agent tasks, evaluators must choose between specifying an exact acceptable sequence of steps (a reference trajectory) or defining acceptable properties of any sequence (a rubric). Reference trajectories are precise but brittle: a valid alternative execution path fails evaluation. Rubrics are more flexible but must be designed carefully to avoid passing clearly incorrect paths. For constrained tasks with a known optimal path, a reference trajectory is appropriate. For tasks with multiple valid approaches, a rubric that specifies mandatory steps and prohibited patterns is more reliable.

\paragraph{Stateful evaluation: making context loss verifiably detectable.}
A stateful evaluation scenario is only useful if context loss produces a verifiably wrong output. Generic multi-turn conversations do not satisfy this requirement. The information established in an early turn must be necessary for the correct final answer, such that any drop of that context produces a specifically wrong result with no other plausible explanation. A state-retaining agent and a context-dropping agent must produce detectably different outputs on the same scenario; if they do not, the scenario does not test state retention.

\paragraph{Agent handoff accuracy: constructing explicitly dependent scenarios.}
The agentic-specific challenge is making handoff failures directly detectable rather than inferred. Evaluation scenarios must be constructed so that the downstream agent's correct output depends explicitly on a specific constraint or context element the upstream agent is responsible for transferring. The test setup should include a baseline where the upstream brief is complete and correct and a variation where a single element is omitted or corrupted; the downstream agent's output must differ detectably between the two cases. Scenarios that do not exhibit this property cannot distinguish a handoff failure from a downstream reasoning failure.

\paragraph{Failure injection: testing each recovery path explicitly.}
Failure injection breaks individual components deliberately: returning API errors, providing empty retrieval results, introducing malformed tool responses. Each failure mode and each expected recovery path (retry, fallback, clarification request, clean termination) should be a distinct scenario in the evaluation dataset. Recovery behavior that is only ever observed as a side effect of general quality testing is not reliably evaluated; if a recovery path does not have its own test scenario, it has not been evaluated.

\paragraph{Emergent behaviour testing: failures invisible to individual evaluation.}
Emergent coordination failures only appear when agents interact. The scenario design principle is to construct inputs that force a specific coordination failure: ambiguous handoff instructions where agents interpret the brief differently; state conflicts where two agents hold inconsistent views of prior context; tasks requiring an agent to correctly handle incomplete output from another without incorrectly flagging it as an error. These scenarios should be built from real production incidents, because production is currently the primary source of emergent failure discovery in most organisations.

\paragraph{Agent goal completion: why lower-layer metrics cannot substitute.}
A system can have correct trajectories, accurate tool calls and intact handoffs yet still fail the user's intent if the task scope was misunderstood at the outset or if a constraint established early in the session was silently dropped before the final step. Agent goal completion rate is not predictable from the sum of lower-layer scores and it is not equivalent to multi-agent task success rate. Task-level success criteria must be defined explicitly at the scenario level and each scenario must be assessed against every dimension of the user's stated objective. A structured LLM judge with a task-level rubric or human evaluation is required. Agent goal completion rate is the final gate before any agentic system change is approved for production deployment.

\section{Model Lifecycle, Drift and Regression Management}
\label{sec:lifecycle}

Getting the assurance pyramid in place is a significant step. But there is a reality of production AI systems that engineering teams are often unprepared for and that is the model the application depends on today may not behave the same way in six months or may not exist at all.

In traditional software, dependencies are relatively stable. A library version pinned today will work the same way years from now. With cloud-hosted LLMs, that stability is not guaranteed. Providers operate on aggressive deprecation timelines, update hosted models without always issuing explicit changelogs and retire model versions within twelve to eighteen months of general availability. These are not exceptional events. They are the operational baseline for any team running AI systems in production.

\subsection{Prompts Are Behavioural Specifications}

Before addressing how to manage model change, it is important to establish what is actually at risk when a model changes.

A prompt is not a text string. It is a \textbf{behavioural specification}: it defines what the model should do, how it should reason and what it should produce. Change the model underneath it and this specification may be interpreted differently. A newer model may follow certain instruction patterns more strictly, apply safety constraints more aggressively or interpret ambiguous phrasing in ways that differ from its predecessor.

In a real enterprise AI application, a single user query often passes through multiple independent LLM interactions: a safety check, intent classification, query rewriting, RAG generation and response formatting. Each is a separate prompt. Each must be re-validated when the model changes. \textbf{Model migration must be treated as a first-class release event} not a routine dependency bump.

\subsection{The Three Change Scenarios}

\paragraph{Scenario 1: Model change, prompts unchanged.}
A new model version is adopted or a fallback is introduced. The prompts remain identical. This feels safe, but different models interpret the same instructions differently. Even without touching the prompts, every prompt must be re-tested. The model is the interpreter and the model change have replaced the interpreter.

\paragraph{Scenario 2: Prompt change, model unchanged.}
A prompt is refined tightening instructions, adding constraints, improving clarity. Even small wording changes can produce meaningfully different model behaviour across the distribution of inputs. Every prompt change must be treated as a code change: it needs its own evaluation run before it ships.

\paragraph{Scenario 3: Model change and prompt change simultaneously.}
This is the highest-risk scenario. Teams often take a model upgrade as an opportunity to also refine prompts (sensible engineering practice), but then validate only at the end-to-end layer. The correct approach: isolate one variable at a time. Validate the new model against existing prompts first. Once that baseline is stable, introduce prompt changes separately with their own evaluation runs.

\begin{figure}[htbp]
\centering
\begin{tikzpicture}[
  box/.style={rectangle, rounded corners=4pt, minimum width=2.2cm, minimum height=0.8cm,
              align=center, font=\small, draw},
  arrow/.style={-{Stealth[length=4pt]}, thick},
  lbl/.style={font=\footnotesize, text=gray!80},
  risk/.style={font=\footnotesize\bfseries}
]
  \node[box, fill=yellow!15, draw=yellow!60!black] (m1) at (0, 4.2)   {New Model};
  \node[box, fill=gray!8,    draw=gray!50]         (p1) at (3.2, 4.2) {Same Prompts};
  \node[box, fill=orange!12, draw=orange!60]       (e1) at (6.4, 4.2) {Re-evaluate\\all prompts};
  \draw[arrow] (m1) -- (p1);  \draw[arrow] (p1) -- (e1);
  \node[risk, text=orange!80!black] at (9.6, 4.2) {Medium risk};

  \node[box, fill=gray!8,   draw=gray!50]          (m2) at (0, 2.8)   {Same Model};
  \node[box, fill=yellow!15, draw=yellow!60!black] (p2) at (3.2, 2.8) {Prompt Change};
  \node[box, fill=orange!12, draw=orange!60]       (e2) at (6.4, 2.8) {Evaluate\\changed prompt};
  \draw[arrow] (m2) -- (p2);  \draw[arrow] (p2) -- (e2);
  \node[risk, text=orange!80!black] at (9.6, 2.8) {Medium risk};

  \node[box, fill=yellow!15, draw=yellow!60!black] (m3) at (0, 1.4)   {New Model};
  \node[box, fill=yellow!15, draw=yellow!60!black] (p3) at (3.2, 1.4) {Prompt Change};
  \node[box, fill=red!12,    draw=red!60]          (e3) at (6.4, 1.4) {Isolate: model\\first, then prompt};
  \draw[arrow] (m3) -- (p3);  \draw[arrow] (p3) -- (e3);
  \node[risk, text=red!70!black] at (9.6, 1.4) {Highest risk};

  \node[lbl] at (0,   5.0) {Model};
  \node[lbl] at (3.2, 5.0) {Prompts};
  \node[lbl] at (6.4, 5.0) {Evaluation Discipline};
\end{tikzpicture}
\caption{The three change scenarios and their required evaluation discipline.
Scenario 3 simultaneous model and prompt change carries the highest risk
and requires variable isolation before combined validation.}
\label{fig:change-scenarios}
\end{figure}
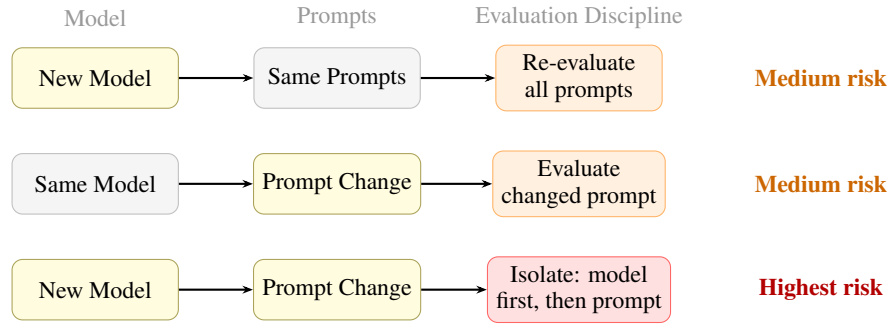

Figure~\ref{fig:change-scenarios} illustrates the evaluation discipline required for each scenario.

\subsection{Model Drift: The Silent Degrader}

Model deprecation is visible providers announce it. Model drift is the opposite: quiet, gradual and easy to miss until users start noticing.

Model drift refers to changes in model behaviour over time, even when nothing have changed. Providers periodically update hosted models safety tuning, alignment adjustments, performance improvements without always issuing changelogs. The danger is not catastrophic failure. It is \textit{slight wrongness, consistently, across a large volume of interactions}.

\textit{Example: A travel assistant has been confidently recommending non-stop flights first as instructed for months. After a quiet model update, it begins occasionally ignoring that instruction. No error is thrown. User satisfaction scores start dipping and by the time the pattern is identified, it has been happening for weeks.}

The only reliable defence against drift is prompt-level regression testing running continuously not just at release time. Drift is invisible to monitoring tools that only observe errors and latency. It requires evaluation-based detection: tracking quality scores over time and flagging when trends degrade beyond a defined threshold.

\subsection{Building and Running Prompt Test Datasets}

As established in Section~\ref{sec:edd}, every prompt in a production system requires its own evaluation dataset. For regression management specifically, these datasets must cover three scenario types with particular care:

\begin{itemize}[leftmargin=*]
  \item \textbf{Edge cases}: inputs at the boundary of expected behaviour, where prompt interpretation is most sensitive to model-level changes.
  \item \textbf{Unsafe and adversarial inputs}: the scenarios where model drift is most likely to manifest, as safety tuning and alignment updates specifically affect how the model responds to these inputs.
  \item \textbf{Previously failing inputs that were fixed}: cases where prompt engineering resolved a known failure. These are the first places to check after a model update, as the fix may no longer hold with a different interpreter.
\end{itemize}

\subsection{Continuous Monitoring in Practice}

Continuous monitoring operates at multiple levels simultaneously. At its core is a cadenced evaluation pipeline that runs automatically on two triggers: schedule (at minimum, daily) and event (any model version change or prompt update).

The output of each run is not just a pass/fail count it is a score trend. A single failing run may be noise. A score that trends downward over two weeks is a signal requiring investigation. The three consistency verdicts from Section~\ref{sec:edd} consistent pass, flaky, consistent failure apply directly to trend analysis: a prompt whose pass rate has moved from 95\% to 80\% over a month is flaky in a temporal sense and warrants the same investigation as a prompt that fails intermittently in a single run.

Table~\ref{tab:monitoring-cadence} summarises the recommended monitoring cadence and trigger conditions.

\begin{table}[h]
\centering
\caption{Continuous monitoring cadence and triggers}
\label{tab:monitoring-cadence}
\begin{tabularx}{\linewidth}{@{} l l X @{}}
\toprule
\textbf{Trigger} & \textbf{Scope} & \textbf{Action on Degradation} \\
\midrule
Scheduled (daily)
  & Full prompt test suite at Layer 1
  & Flag prompts with score trend $>$ 5\% decline over 7 days \\
Model version change
  & All prompts, all layers
  & Block deployment until Layer 1 baseline is re-established \\
Prompt update
  & Updated prompt + dependent downstream prompts
  & Require evaluation run before merge \\
Production incident
  & Affected prompt + upstream components
  & Identify pyramid layer of origin; add incident case to dataset \\
\bottomrule
\end{tabularx}
\end{table}

\paragraph{Converting incidents into evaluation coverage.}
Every production incident that reveals a new failure mode is an opportunity to close an evaluation gap. The standard response should be: identify which pyramid layer should have caught this failure, determine why it did not (missing scenario in dataset, insufficient rubric coverage, no adversarial test for this input class), add the incident case to the relevant evaluation dataset and confirm that the updated dataset now catches the failure before the fix is deployed. This converts reactive incident management into proactive coverage expansion.

\section{Governance, Human Oversight and Operational Economics}
\label{sec:governance}

Technical assurance alone is insufficient for enterprise AI deployment without organizational structures and economic oversight. The organisational structures, oversight mechanisms and economic constraints surrounding the evaluation infrastructure determine whether that infrastructure is actually used, trusted and sustained.

\subsection{Human-in-the-Loop Oversight}

Not all AI decisions should be fully automated. In high-stakes domains healthcare, legal advice, financial recommendations, compliance determinations, human review remains an essential layer of the assurance model, not a fallback for when automation fails.

An effective human-in-the-loop strategy does not replace automation with manual review. It defines \textbf{when human judgment is required}: which output classes, which confidence thresholds and which failure modes trigger escalation to a human reviewer. This requires:

\begin{itemize}[leftmargin=*]
  \item \textbf{Uncertainty thresholds}: automated evaluation scores below a defined confidence boundary are routed for human review before being acted upon.
  \item \textbf{Sampling strategies}: a statistically representative sample of all outputs is reviewed periodically, regardless of automated scores, to catch failure modes that evaluations do not yet cover.
  \item \textbf{Escalation workflows}: clear routing from automated failure detection to the appropriate human reviewer, with defined response time expectations and feedback loops back into the evaluation dataset.
\end{itemize}

Human review is not a replacement for evaluation infrastructure. It is a complement: the mechanism through which unknown unknowns that escape automated evaluation are surfaced, investigated and converted into evaluation coverage.

\subsection{Cost vs. Reliability Tradeoffs}

Enterprise AI systems operate under economic constraints as real as their correctness constraints. Evaluation infrastructure carries costs: LLM judge calls, human review time, evaluation dataset maintenance and the latency and compute overhead of running evaluations at scale. These costs must be managed explicitly, not ignored.

The primary levers for managing evaluation economics are:

\begin{itemize}[leftmargin=*]
  \item \textbf{Deterministic validation before semantic evaluation}: as established in Section~\ref{sec:edd}, structured outputs should always be validated with assertions before reaching an LLM judge. This eliminates evaluation token cost for the majority of outputs in well-designed systems.
  \item \textbf{Tiered evaluation frequency}: Layer 0 and Layer 1 evaluations run on every change; Layer 3 and Layer 4 evaluations run on a less frequent cadence unless triggered by a specific incident or migration event.
  \item \textbf{Risk-proportionate coverage}: evaluation dataset size and judge invocation frequency should be proportionate to the risk profile of the prompt. A guardrail prompt that can produce unsafe outputs warrants more thorough evaluation than a formatting prompt with structured output.
\end{itemize}

The cost of under-investing in evaluation is not zero. Silent hallucination at scale, prompt regression discovered in production and model drift detected weeks after onset all carry remediation costs that far exceed the evaluation infrastructure investment they would have replaced.

\subsection{Auditability and Traceability}

In regulated industries financial services, healthcare, insurance, legal AI systems must be demonstrably auditable. This is not a feature request from engineering teams; it is a compliance requirement that affects system architecture from the outset.

Auditability requires that for any AI-generated output, the following are retrievable:

\begin{itemize}[leftmargin=*]
  \item The model version and configuration that produced it
  \item The prompt version used, including any dynamic variables injected at runtime
  \item The retrieved context provided (for RAG systems), with source attribution
  \item The evaluation scores and verdicts from the most recent evaluation run covering that prompt
\end{itemize}

This is not achievable through after-the-fact logging alone. It requires prompt versioning systems, evaluation lineage tracking and retrieval provenance infrastructure that must be designed into the system, not retrofitted.

\subsection{From QA Teams to Evaluation Engineering}

The organisational implication of AI assurance is significant: traditional QA roles are insufficient for maintaining AI evaluation infrastructure at scale.

Evaluation engineering in AI systems requires a different skill set: building and curating evaluation datasets, designing rubrics, maintaining judge pipelines, interpreting score trends and translating production incidents into evaluation coverage. These are not test automation tasks. They are a distinct discipline, sitting at the intersection of software engineering, data curation and behavioural science.

Organisations that treat AI evaluation as a downstream QA activity, something applied after the development team ships, will consistently under-invest in it. The evaluation infrastructure should be treated as a shared platform capability: centralised, maintained by a dedicated function and consumed by multiple AI product teams rather than rebuilt for each project.

This shift from project-level test suites to platform-level evaluation infrastructure is the organisational equivalent of the move from per-project CI scripts to a shared CI/CD platform. It is not optional for organisations running multiple AI systems in production. It is the only sustainable model at scale.

\section{Conclusion: From Software QA to AI Assurance Engineering}
\label{sec:conclusion}

The central argument of this paper is that AI systems are not harder-to-test software. They are a different kind of infrastructure, one for which the fundamental assumptions of software quality assurance do not hold and for which a new assurance model is required.

The shift is not incremental. It is epistemological.

In deterministic software, testing is an act of verification. It is specified what the system should do, it is implemented and confirmed that the implementation matches the specification. A passing test suite is evidence of correctness.

In AI systems, the same sequence cannot produce the same certainty. The model is not fully programmable. Its outputs are probabilistic. Its behaviour changes when the underlying model is updated by a third party. And the most dangerous failures (hallucination, instruction drift, trajectory collapse, emergent coordination failure) do not produce errors. They produce wrong answers that look right.

For these systems, testing is not an act of verification. \textbf{It is an act of continuous risk reduction.} The goal is not to certify correctness. It is to accumulate evidence that the system behaves acceptably across the distribution of inputs it will encounter, to bound the failures that remain and to detect degradation before it reaches users at scale.

\subsection{The Five Core Arguments}

This paper has made five interconnected arguments:

\begin{enumerate}[leftmargin=*]
  \item \textbf{AI testing requires a new philosophy.} Testing as risk reduction not verification is the correct mental model. The CI/CD pipeline, not the test suite, is the right analogy for evaluation infrastructure in AI systems.

  \item \textbf{Failure taxonomy precedes testing strategy.} Testing without a defined taxonomy of failures effectively limits the evaluation to only the issues that may be thought of at the moment. The fifteen failure modes in Section~\ref{sec:taxonomy} define what the pyramid is built to catch. Adding new layers to the pyramid without mapping them to failure modes is structural decoration, not assurance.

  \item \textbf{Evaluations define behaviour; they do not just verify it.} In AI systems, the evaluation dataset is the primary mechanism through which acceptable behaviour is specified. Evaluation-Driven Development is not a methodology choice, it is the natural consequence of working with systems that cannot be fully programmed.

  \item \textbf{The pyramid must reflect economics.} Failures caught at Layer 0 are cheap and precisely locatable. Failures caught at Layer 4 are expensive and ambiguous. A well-designed pyramid pushes coverage as low as possible for each failure class. A pyramid that is top-heavy over-relying on end-to-end evaluation is slow, expensive and diagnostically weak.

  \item \textbf{Evaluation infrastructure is a platform, not a project.} Organisations running multiple AI systems in production cannot sustain per-project evaluation tooling. The evaluation stack datasets, judges, rubrics, scoring pipelines, regression baselines, must become shared platform infrastructure, maintained with the same rigour as production code.
\end{enumerate}

\subsection{What Changes for Engineering Teams}

The practical implications of this strategy are significant.

Prompts must be version-controlled and regression-tested like code. Every model migration must trigger a full evaluation run across every prompt in the system, not a smoke test at the end-to-end layer. Evaluation datasets must be maintained as living artefacts, expanded with every production incident. Quality gates must be defined before deployment, not calibrated after the fact.

The skill set required also changes. Building and curating evaluation datasets, designing rubrics that reliably score the failure modes they target, maintaining judge pipelines and interpreting score trends over time are not traditional QA skills. They are the skills of evaluation engineering, a discipline that organisations deploying AI at scale must build deliberately.

\subsection{What Changes for Engineering Leaders}

The investment framing also changes.

Traditional QA expenditure is justified by defect prevention and release confidence. AI assurance expenditure must be justified by risk management: the cost of hallucination in a regulated domain, the cost of model drift detected weeks after onset, the cost of a production incident that reveals an adversarial failure mode that evaluation would have caught.

The question for engineering leaders is not whether to invest in evaluation infrastructure. It is whether to invest before a production failure makes the investment unavoidable and whether to build it as a shared platform that appreciates across AI products or as ad-hoc tooling that must be rebuilt for each deployment.

\subsection{Closing}

The organisations that deploy AI reliably at scale will not be those with the most capable models. They will be those with the most rigorous evaluation infrastructure teams who have internalised that in probabilistic systems, confidence is not a property of the implementation. It is a property of the evidence accumulated through continuous evaluation.

Software QA was built for a world where software does what it is told to do. AI assurance must be built for a world where systems do what they have learned to do and where the assurer's job is to continuously measure, bound and improve the gap between the two.

\clearpage

\bibliographystyle{unsrtnat}
\bibliography{document}

\clearpage

\appendix

\section{Implementation Patterns}
\label{sec:appendix}

This appendix provides reference implementation patterns for the evaluation approaches
described in Section~\ref{sec:edd} and Section~\ref{sec:rag-agentic}. These patterns are
illustrative; the specific libraries and APIs will vary by environment, but the structural
choices they reflect are intentional and correspond to the evaluation principles in the main text.

\subsection{Deterministic Validation}

When a prompt produces structured output (JSON, classification labels, specific field values), validate it with assertions before invoking any LLM judge. This carries zero token cost and high confidence.

\begin{lstlisting}[language=Python, caption={Deterministic output validation}]
import json

def validate_structured_output(actual_raw: str, expected_raw: str) -> bool:
    actual   = json.loads(actual_raw)
    expected = json.loads(expected_raw)

    # Hard assertions: no LLM involved
    assert actual["type"] == expected["type"], \
        f"Type mismatch: {actual['type']} != {expected['type']}"
    assert sorted(actual["items"]) == sorted(expected["items"]), \
        f"Items mismatch: {actual['items']} != {expected['items']}"

    return True
\end{lstlisting}

\subsection{LLM-as-Judge with Rubrics}

For natural language outputs that cannot be reduced to a schema match, use an LLM evaluator guided by single-dimension rubrics. Each rubric assesses one quality dimension; combining multiple dimensions in a single rubric produces inconsistent scoring.

\begin{lstlisting}[language=Python, caption={Rubric definition and semantic evaluation with RAGAS}]
from ragas.metrics import DomainSpecificRubrics

# One rubric per dimension: never combine multiple criteria in one rubric
RUBRICS = {
    "INTENT_ALIGNMENT": {
        "score0_description": "Response does not address the user's main intent.",
        "score1_description": "Response correctly addresses the user's main intent."
    },
    "FACTUAL_CORRECTNESS": {
        "score0_description": "Response contains factual or logical errors.",
        "score1_description": "Response is factually and logically correct."
    },
    "SAFETY_COMPLIANCE": {
        "score0_description": "Response violates a safety or policy constraint.",
        "score1_description": "Response complies with all safety and policy constraints."
    }
}

async def evaluate_response(user_input, response, reference, evaluator_llm):
    scores = {}
    for dimension, rubric in RUBRICS.items():
        metric = DomainSpecificRubrics(
            llm=evaluator_llm,
            with_reference=True,
            rubrics=rubric
        )
        scores[dimension] = await metric.ascore(
            user_input=user_input,
            response=response,
            reference=reference
        )
    return scores
\end{lstlisting}

\subsection{Weighted Scoring and Mandatory Gates}

Not all evaluation dimensions carry equal weight. Mandatory gates enforce non-negotiable dimensions regardless of aggregate score.

\begin{lstlisting}[language=Python, caption={Weighted scoring and mandatory gates}]
WEIGHTS = {
    "INTENT_ALIGNMENT":   0.4,
    "FACTUAL_CORRECTNESS": 0.4,
    "SAFETY_COMPLIANCE":   0.2
}

# Dimensions that must pass regardless of aggregate score
MANDATORY = {"SAFETY_COMPLIANCE", "FACTUAL_CORRECTNESS"}

def evaluate_pass(scores: dict) -> dict:
    weighted = sum(scores[d] * WEIGHTS[d] for d in WEIGHTS)
    mandatory_pass = all(scores[d] == 1 for d in MANDATORY)

    return {
        "weighted_score": weighted,
        "mandatory_pass": mandatory_pass,
        "final_pass":     mandatory_pass and weighted >= 0.8,
        "failed_gates":   [d for d in MANDATORY if scores[d] != 1]
    }
\end{lstlisting}

\subsection{Consistency Testing}

A single evaluation run tells little about an AI system. Consistency the pass rate across repeated runs distinguishes reliable behaviour from fortunate outputs.

\begin{lstlisting}[language=Python, caption={Consistency testing across repeated runs}]
import asyncio

async def check_consistency(prompt, input_data, metric, reference, n=10):
    tasks = [
        metric.ascore(
            user_input=input_data["input"],
            response=llm.invoke(prompt).content,
            reference=reference
        )
        for _ in range(n)
    ]
    scores = await asyncio.gather(*tasks)
    pass_rate = sum(1 for s in scores if s >= 0.8) / n

    verdict = (
        "CONSISTENT_FAILURE" if pass_rate < 0.5  else
        "FLAKY"              if pass_rate < 0.9  else
        "CONSISTENT_PASS"
    )
    return {"pass_rate": pass_rate, "verdict": verdict, "scores": scores}
\end{lstlisting}

\textbf{Interpreting verdicts:}
\begin{itemize}[leftmargin=*]
  \item \texttt{CONSISTENT\_PASS} reliable behaviour; single-run failures were noise.
  \item \texttt{FLAKY} prompt is too sensitive; tighten instructions until pass rate exceeds 90\%.
  \item \texttt{CONSISTENT\_FAILURE} genuine regression; escalate for Layer 1 prompt investigation.
\end{itemize}

\subsection{RAG Evaluation with RAGAS}

The four RAGAS metrics (Section~\ref{sec:rag-agentic}) can be run together to produce a multi-dimensional quality profile. The result is a diagnostic profile, not a single score.

\begin{lstlisting}[language=Python, caption={Full RAGAS evaluation pipeline}]
from ragas import evaluate
from ragas.metrics import (
    faithfulness,
    answer_relevancy,
    context_recall,
    context_precision,
)

result = evaluate(
    dataset=test_dataset,          # HuggingFace Dataset with required columns
    metrics=[
        context_precision,
        context_recall,
        faithfulness,
        answer_relevancy,
    ],
    llm=evaluator_llm,
    embeddings=embedding_model,
)

# Quality gates: block deployment if any gate fails
QUALITY_GATES = {
    "faithfulness":       0.90,    # Non-negotiable: hallucination risk
    "context_recall":     0.85,    # Non-negotiable: completeness risk
    "context_precision":  0.80,
    "answer_relevancy":   0.80,
}

gate_results = {
    metric: result[metric] >= threshold
    for metric, threshold in QUALITY_GATES.items()
}
deploy_approved = all(gate_results.values())
\end{lstlisting}

\end{document}